\documentclass[11pt,amsmath,aps,amssymb,superscriptaddress,notitlepage,onecolumn,longbibliography]{revtex4-1}
\usepackage{amsmath}
\usepackage{enumitem}
\usepackage{amssymb}
\usepackage{overpic}
\usepackage{soul}
\usepackage[linktoc=all]{hyperref}
\usepackage{float,graphicx}
\usepackage[usenames, dvipsnames]{color}

\hypersetup{
	colorlinks,
	citecolor=blue,
	filecolor=red,
	linkcolor=black,
	urlcolor=blue
}

\usepackage[page]{appendix}

\newcommand{\jump}[1]{\ensuremath{[\![#1]\!]} }

\newcommand{\bu}{{\mathbf u}}

\newcommand{\bp}{{\mathbf p}}
\newcommand{\bx}{{\mathbf x}}

\begin{document}


\title{Transport and dispersion of active particles in periodic porous media}

\author{Roberto Alonso-Matilla}
\altaffiliation{Present address: Department of Chemical Engineering, Columbia University, New York, NY 10027. }
\author{Brato Chakrabarti}
\author{David Saintillan}
\email[Email address for correspondence: ]{dstn@ucsd.edu}
\affiliation{Department of Mechanical and Aerospace Engineering, University of California San Diego, 9500 Gilman Drive, La Jolla, CA 92093, USA}


\date{\today}

\begin{abstract}

The transport of self-propelled particles such as bacteria and phoretic swimmers through crowded heterogeneous environments is relevant to many natural and engineering processes, from biofilm formation and contamination processes to transport in soils and biomedical devices. While there has been experimental progress, a theoretical understanding of mean transport properties in these systems has been lacking. In this work, we apply generalized Taylor dispersion theory to analyze the long-time statistics of an active self-propelled Brownian particle transported under an applied flow through the interstices of a periodic lattice that serves as an idealization of a porous medium.\ Our theoretical model, which we validate against Brownian dynamics simulations, is applied to unravel the roles of motility, fluid flow, and lattice geometry on asymptotic mean velocity and dispersivity. In weak flows, transport is dominated by active dispersion, which results from self-propulsion in the presence of noise and is hindered by the obstacles that act as entropic barriers. In strong flows, shear-induced Taylor dispersion becomes the dominant mechanism for spreading, with pillars now acting as regions of shear production that enhance dispersion.\ The interplay of these two effects  leads to complex and unexpected trends, such as a non-monotonic dependence of axial dispersivity on flow strength and a reduction in dispersion due to swimming activity in strong flows.\ Brownian dynamics are used to cast light on the pre-asymptotic regime, where tailed distributions are observed in agreement with recent experiments on motile micro-organisms. Our results also highlight the subtle effects of pillar shape, which can be used to control the magnitude of dispersion and to drive a net particle migration in quiescent systems.

	
\end{abstract}

\maketitle

\section{Introduction}

The transport of active self-propelled particles through complex microstructured media has important implications in microbial ecology as well as human health.\ Examples include the spreading of contaminants in soils and groundwater aquifers, bacterial filtering, biodegradation and bioremediation processes, and the transport of motile cells inside the body.\ Engineering applications such as medical diagnostics and biochemical analysis also often hinge on the manipulation and control of active particle motions through complex geometries \cite{bechinger2016active}.

While the effective transport of passive tracers in porous media flows has been analyzed in great detail in the past, the case of active particles that self-propel remains largely unexplored.\ In free space, a microswimmer with constant speed $v_0$ undergoing  translational and rotational diffusion with diffusivities $d_t$ and $d_r$ performs a spatial random walk with long-time dispersivity  {$\overline{D}=d_t+v_0^2/2nd_r$}  in $n$ dimensions \cite{berg1993random}.\ The dispersion enhancement due to swimming, henceforth termed \textit{active dispersion}, is the combined effect of rotational diffusion (which allows the particle to sample all swimming directions) and advective transport by swimming, resulting in a correlated random walk in space.\ In a porous matrix and in the absence of flow, this {active} dispersion should intuitively be reduced, as indeed seen in experiments on \textit{Chlamydomonas reinhardtii} in pillar arrays \cite{Brun2018}.\ This reduction is due in part to geometric obstruction, with the pillars acting as entropic barriers \cite{laachi2007}, but also to the natural tendency of microswimmers to accumulate at boundaries \cite{ezhilan2015channel,ezhilan2015b,yan2015boundary}.\ Trapping near obstacles is indeed seen in experiments \cite{takagi2014hydrodynamic,creppy2018motility} where it is also thought to be enhanced by swimmer-wall hydrodynamic interactions \cite{spagnolie2015geometric}.

When an external flow is applied, the coupling of particle orientations with the fluid shear further complicates transport.\ First, diffusion and swimming across streamlines alter long-time spreading by the classic phenomenon of \textit{shear-induced dispersion}. This dispersion mechanism,  first analyzed by Taylor \cite{taylor1953dispersion} and Aris \cite{aris1956dispersion} in the context of passive tracers in pressure-driven pipe flow, results from the random sampling of fluid velocities by the particles as they stochastically travel across velocity gradients, causing them to disperse in the streamwise direction. Secondly, the rotation of particles in the strong shear near boundaries is expected to result in their alignment against the flow \cite{altshuler2013flow,creppy2018motility}, possibly leading to a net reduction in mean transport by an effect similar to upstream swimming in pressure-driven flows \cite{kaya2012upstream,ezhilan2015channel}. Recent experiments on \textit{Escherichia coli} in random post arrays seem to support this hypothesis \cite{creppy2018motility}.\ The situation is yet more complex in the presence of asymmetric obstacles, where the gliding of microswimmers along curved boundaries can drive a net migration even in quiescent conditions \cite{galajda2007wall,yariv2014,wykes2017guiding,tong2017directed}. 

Some of these trends have been confirmed in experiments \cite{brown2016swimming,sosa2017motility,morin2017diffusion,creppy2018motility,Brun2018} as well as various  computational models \cite{chepizhko2013diffusion,potiguar2014self,chamolly2017active,khalilian2016obstruction}, yet a first-principles theory for predicting the statistics of active transport in even simple porous matrices remains lacking. Transport and dispersion of active particles has previously been analyzed in simpler settings such as in unbounded linear flows \cite{hill2002gyrolinear,manela2003gyroshear,sandoval2014linear} as well as in straight \cite{bees2010gyrotube,Chilukuri2015channel} and corrugated \cite{yariv2014,sandoval2014channel} channels.\ Several of these studies have been based on generalized Taylor dispersion theory (GTDT), a theoretical method first developed by Brenner and coworkers \cite{brenner1980pch,frankel1989taylordisp,brenner2013macrotransport} to describe shear-induced dispersion.\ More fundamentally, GTDT (also known as macrotransport theory \cite{brenner2013macrotransport}) provides a general framework for calculcating the long-time statistics of particle configurations under the coupled effects of advective transport (e.g., fluid flow, swimming, sedimentation, chemotaxis) and diffusive transport (e.g., translational or rotational diffusion, run-and-tumble dynamics). Both active and shear-induced dispersion as well as upstream swimming fall within the purview of GTDT, which provides a convenient tool for the theoretical analysis of these effects and of their interplay in porous media flows of self-propelled particles.  

In this work, we present a theoretical model based on GTDT for the transport and effective spreading of microswimmers in two-dimensional periodic lattices.\ The theory extends previous studies on transport of passive point tracers in porous media \cite{brenner1980dispersion,edwards1991dispersion} to the case of orientable particles \cite{brenner1979} that can also swim along their unit director.\ We adopt in Sec.~\ref{sec:problemdef} a very simple physical model of an active Brownian particle that self-propels, is transported and rotated by an applied flow, and diffuses in both space and orientation.\ To highlight the universal features of active transport, we only retain these key ingredients, as they are common to most types of self-propelled particles from motile micro-organisms to phoretic swimmers. In particular, we neglect higher-order swimmer-specific effects such as hydrodynamic interactions, which may become significant in some experimental realizations.\ We also focus on two-dimensional geometries, which are more tractable and also very relevant to describe experimental systems  using microfabricated pillar arrays in Hele-Shaw channels \cite{creppy2018motility,Brun2018}.\ Starting from the Fokker-Planck equation for the configuration of a single particle in a periodic lattice  in Sec.~\ref{sec:theory}, we apply Brenner's method of moments \cite{brenner1980pch,brenner1980dispersion} to derive theoretical expressions for the asymptotic transport velocity and dispersivity dyadic, 
 \begin{equation}
 \overline{\mathbf{U}}=\lim_{t\rightarrow\infty} \frac{d}{dt}\langle \mathbf{R}(t)\rangle, \qquad \overline{\mathbf{D}}=\lim_{t\rightarrow\infty}\frac{1}{2}\frac{d}{dt}\langle \left(\mathbf{R}(t) - \langle \mathbf{R}(t) \rangle \right) \left(\mathbf{R}(t) - \langle \mathbf{R}(t) \rangle \right)\rangle,
 \end{equation} 
where $\mathbf{R}(t)$ is the instantaneous swimmer position and $\langle\cdot\rangle$ denotes the ensemble average. Results from our theory are tested against Brownian dynamics simulations in Sec.~\ref{sec:results}, where we unravel the roles of particle motility, fluid flow, and lattice geometry on long-time transport properties. We summarize our findings and conclude in Sec.~\ref{sec:conclusions}.

\section{Problem Definition and Langevin description\label{sec:problemdef}}

\begin{figure}[t]
	\centering
	\includegraphics[width=0.85\linewidth]{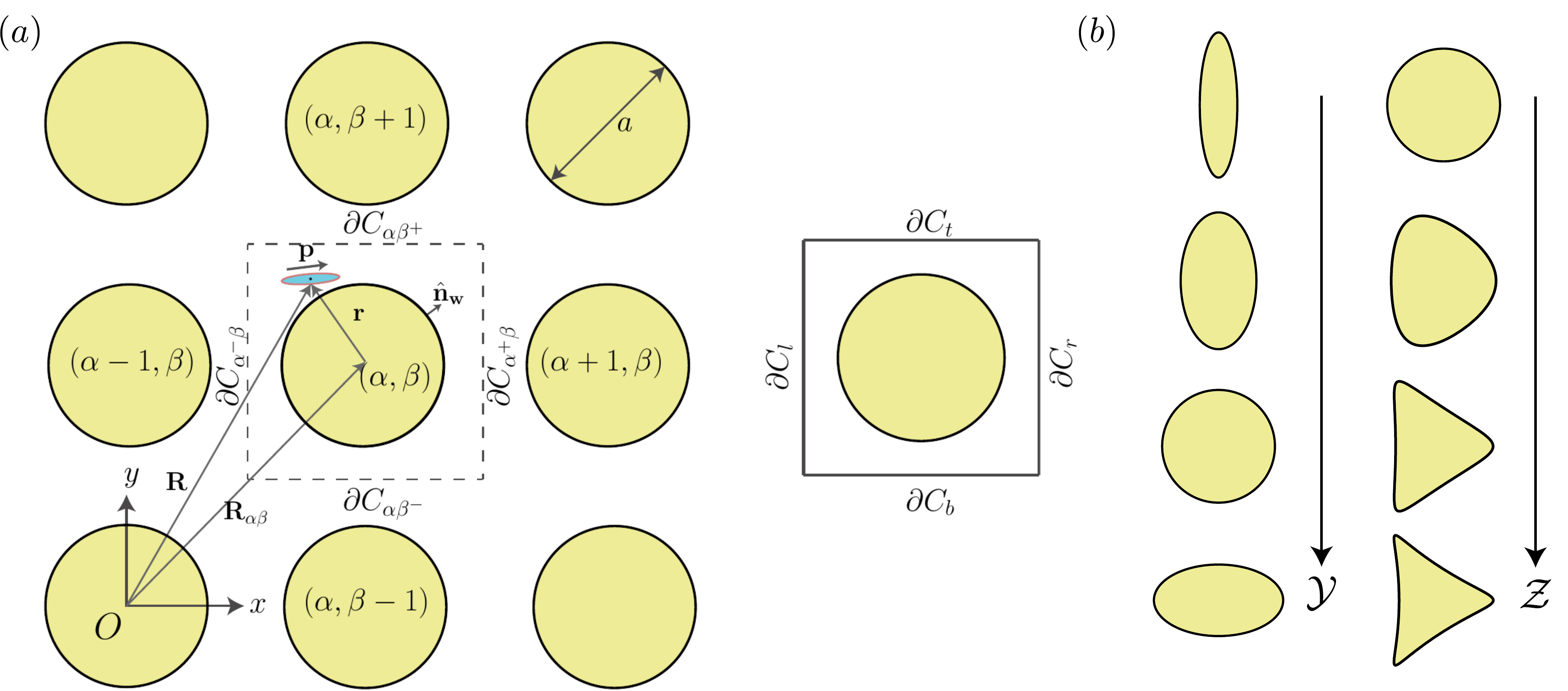}\vspace{-0.2cm}
	\caption{($a$) Geometry of a representative lattice, unit cell and associated nomenclature. A typical unit cell for a square lattice is shown on the right. ($b$) Non-spherical obstacle shapes are generated by an area-preserving conformal mapping \cite{avron2004optimal}, where parameters $\mathcal{Y}$ and $\mathcal{Z}$ control obstacle aspect ratio and fore-aft asymmetry, respectively (see Appendix~\ref{app:CM}).  } \label{fig:geo} 
\end{figure}

We analyze the long-time asymptotic transport of a dilute collection of self-propelled active particles of spherical shape dispersed in a viscous solvent and moving through the interstices of a doubly-periodic two-dimensional porous material.\ The porous medium is modeled as an infinite lattice comprised of rigid obstacles of characteristic dimension $a$ and generated through the discrete translation of a geometric unit cell of linear dimensions~$L_x\times L_y$ (Fig.~\ref{fig:geo}).\ Each cell in the lattice is labeled by two integers $(\alpha,\beta) \in\mathbb{Z}^2$, which identify its position $\mathbf{R}_{\alpha\beta}=(X_{\alpha},Y_{\beta})=(\alpha L_x,\beta L_y)$ with respect to origin $O$, chosen as the centroid of an arbitrary cell.\ The array is characterized by its aspect ratio $\delta=L_y/L_x$ and by its porosity $\epsilon_p=S_f/S_t$, or ratio of the interstitial fluid area $S_f$ of a cell over its total area $S_t$. While the focus of the paper is on square lattices of circular pillars as shown in Fig.~\ref{fig:geo}($a$), we also present a few results on the effects of lattice arrangement and pillar shape in Sec.~\ref{sec:latticeproperties}. The pillar shapes we consider in those examples are illustrated in Fig.~\ref{fig:geo}($b$) and are obtained by conformal mapping of the unit circle \cite{avron2004optimal}. The mapping we employ preserves pillar area and involves two parameters $\{\mathcal{Y},\mathcal{Z}\}$ that control pillar aspect ratio and fore-aft asymmetry, respectively (see Appendix~\ref{app:CM} for details).

In a dilute system, we can neglect interparticle interactions and need only consider the transport of a single swimmer.\ Its instantaneous configuration is described by its position $\mathbf{R}(t)$ with respect to point $O$ and by its swimming direction $\mathbf{p}=(\cos{\theta},\sin{\theta})$ in the plane of motion, with polar angle $\theta(t)\in\Omega\equiv[0,2 \pi)$.\ The global coordinate $\mathbf{R}$ can be decomposed as $\mathbf{R}=\mathbf{R}_{\alpha\beta}+\mathbf{r}$, where $\mathbf{R}_{\alpha\beta}$ is the position of the unit cell where the particle is located and $\mathbf{r}=(x,y)$ is a local coordinate with respect to the center of that cell.\ The particle is released at $t=0$ at position $\mathbf{R}_{0}=\mathbf{r}_0$ with orientation $\mathbf{p}_0$.\ We introduce the following gradient operators:\ global $\nabla_R \equiv \partial/\partial \mathbf{R}$, local $\smash{\nabla_r \equiv \left(\partial/\partial \mathbf{r}\right)_{\mathbf{R_{\alpha \beta}}}}$ and orientational $\smash{\nabla_p \equiv  (\mathbf{I}-\mathbf{pp})\cdot\partial/\partial \mathbf{p}}$. 

We use the simple model of an active Brownian particle, in which particle motion results from self-propulsion, from advection and rotation by the fluid flow, and from translational and rotational diffusion.\  Particle dynamics are described by two coupled Langevin equations: 
\begin{align}
\dot{\mathbf{R}}&=v_0\mathbf{p}+\mathbf{u}(\mathbf{r})+\boldsymbol{\eta}_t(t)\,, \label{eq:xdot} \\
\dot{\mathbf{p}}\,&=\frac{1}{2}\boldsymbol{\Omega}(\mathbf{r})\times \mathbf{p}+(\mathbf{I}-\mathbf{pp})\cdot\boldsymbol{\eta}_r(t)\,, \label{eq:pdot}  
\end{align}
where $\boldsymbol{\eta}_t$ and $\boldsymbol{\eta}_r$ are Gaussian random vectors with zero mean that satisfy the fluctuation-dissipation theorem \cite{chaikin1995}:\ $\langle \boldsymbol{\eta}_{t}(t)\boldsymbol{\eta}_t(t')\rangle =\sqrt{2d_t}\,\mathbf{I}\,\delta(t-t')$ and $\langle \boldsymbol{\eta}_{r}(t)\boldsymbol{\eta}_r(t')\rangle =\sqrt{2d_r}\,(\mathbf{I}-\mathbf{pp})\,\delta(t-t')$, respectively.\ The velocity field $\mathbf{u}(\mathbf{r})$ is a two-dimensional Stokes flow driven by a macroscopic pressure gradient applied across the array, with uniform speed $u_\infty$ and incoming angle $\Theta_f$ upstream of the array; we solve for it numerically by the boundary integral method (see Appendix~\ref{app:BI} for details).\ The fluid vorticity $\mathbf{\Omega}(\mathbf{r})=\nabla_r\times \mathbf{u}$ also causes rotation of the particle in Eq.~(\ref{eq:pdot}), which assumes a spherical shape \cite{jeffery1922motion}.\ Our aim is to describe the long-time statistics of the particle position $\mathbf{R}(t)$, which we model within the continuum framework of generalized Taylor dispersion theory and validate against Brownian dynamics simulations.

\section{Continuum Model\label{sec:theory}}

\subsection{Fokker-Planck description}

We introduce the conditional probability density $\psi(\mathbf{R},\mathbf{p},t|\mathbf{R}_0,\mathbf{p}_0)$ of finding the particle at a position $\mathbf{R}$ with orientation $\mathbf{p}$ at time $t$, given that it was initially released at $(\mathbf{R}_0,\mathbf{p}_0)$.\ It satisfies the Fokker-Planck equation \cite{chandrasekhar1943stochastic,doiedwards1986} 
\begin{equation}
\frac{\partial \psi}{\partial t}+\nabla_R \cdot \mathbf{J}+\nabla_p \cdot \mathbf{j}=\delta(\mathbf{R}-\mathbf{R}_0)\delta(\mathbf{p}-\mathbf{p}_0) \delta(t)\,, \label{eq:FP} \vspace{-0.15cm}
\end{equation}
where {the initial condition} is incorporated by the source term on the right-hand side, and where we have assumed $\psi = 0$ for $t<0$.  We henceforth use dimensionless equations where we scale variables using length scale $a$, time scale $d_r^{-1}$ and  velocity scale $ad_r$, and we also normalize $\psi$ by $1/a^2$.\ In addition to $\delta$ and $\smash{\epsilon_p}$ introduced in Sec.~\ref{sec:problemdef}, non-dimensionalization of the governing equations yields three dimensionless groups \cite{ezhilan2015channel}: 
\begin{equation}
\mathrm{Pe}_s=\frac{v_0}{a d_r}, \qquad \mathrm{Pe}_f=\frac{u_{\infty}}{a d_r}, \qquad \kappa=\sqrt{\frac{d_t}{a^2 d_r}}.
\end{equation}
The swimming P\'eclet number $\mathrm{Pe}_s$ is the ratio of the characteristic time scale for a particle to lose memory of its orientation due to rotational diffusion over the time it takes it to swim across the characteristic size of an obstacle. The flow P\'eclet number $\mathrm{Pe}_f=u_{\infty}/a d_r$ compares the characteristic diffusive time to the time scale of alignment under the imposed shear and serves as a measure of flow strength. Finally, $\kappa$ is a fixed constant relating the translational and rotational diffusivities of the particle.  The dimensionless spatial and orientational probability fluxes are easily inferred from the Langevin equations (\ref{eq:xdot})--(\ref{eq:pdot}) as:
\begin{align}
\mathbf{J}&=\left[\mathrm{Pe}_s \mathbf{p}+ \mathrm{Pe}_f \mathbf{u}(\mathbf{r})\right]\psi-\kappa^2 \nabla_R {\psi}, \label{eq:x_flux}  \\
\mathbf{j}\,&= \frac{1}{2}\mathrm{Pe}_f \left[\boldsymbol{\Omega}(\mathbf{r})\times \mathbf{p}\right]\psi-(\mathbf{I}-\mathbf{pp})\cdot\nabla_p {\psi}.\label{eq:p_flux} 
\end{align}
In a given cell ($\alpha,\beta$), Eq.~(\ref{eq:FP}) can be recast solely in terms of local variables as
\begin{equation}
\frac{\partial \psi}{\partial t}+\nabla_r \cdot \mathbf{J}+\nabla_p \cdot\mathbf{j}=\delta_{\alpha 0}\delta_{\beta 0}\delta(\mathbf{r}-\mathbf{r}_0)\delta(\mathbf{p}-\mathbf{p}_0) \delta(t), \label{eq:FP_local}
\end{equation}
where $\psi \equiv \psi(\mathbf{R}_{\alpha \beta},\mathbf{r},\mathbf{p},t| \mathbf{r}_0,\mathbf{p}_0)$. It is normalized in such a way that the sum of the probabilities over all possible configurations is unity after the swimmer is set free:
\begin{equation}
\sum_{\alpha,\beta=-\infty}^{\infty} \int_{S_{\alpha\beta}} \int_{\Omega}  \psi\,  d\mathbf{p}\, d^2\mathbf{r}=\left\{
                \begin{array}{ll}
                  1 \quad \text{for} \quad t \ge 0\,,\\
                  0 \quad \text{for} \quad t < 0\,.
                \end{array}
              \right. \label{eq:norma_cond} 
\end{equation}
Convergence of the previous infinite sum requires that
\begin{equation}
\psi(\mathbf{R_{\alpha\beta}},\mathbf{r},\mathbf{p},t| \mathbf{r}_0,\mathbf{p}_0) \to 0 \quad \text{as} \quad \alpha,\beta \to \pm \infty. \label{eq:convergence1}
\end{equation}
Boundary conditions for $\psi$ need to be specified on the obstacle boundaries as well as cell edges. The no-flux condition at the obstacle walls reads
\begin{equation}
\mathbf{n}_w \cdot \mathbf{J} = 0 \quad \text{on} \quad \partial C_w, \label{eq:no-flux}
\end{equation}
where $\mathbf{n}_w$ is the outward unit vector normal to the wall. Continuity across the cell edges requires
\begin{align}
	&\psi(\mathbf{R_{\alpha\beta}},\mathbf{r},\mathbf{p},t)= \psi(\mathbf{R}_{\alpha+1\beta},\mathbf{r}-L_x\mathbf{e}_x,\mathbf{p},t)  \quad \text{on} \quad \mathbf{r} \in \partial C_{\alpha^+\beta}, \label{eq:BC_psi_edgesip}\\
	&\psi(\mathbf{R_{\alpha\beta}},\mathbf{r},\mathbf{p},t)= \psi(\mathbf{R}_{\alpha\beta+1},\mathbf{r}-L_y\mathbf{e}_y,\mathbf{p},t)  \quad \text{on} \quad \mathbf{r} \in \partial C_{\alpha\beta^+}, \label{eq:BC_psi_edgesjp}
\end{align}
where we omit arguments $\mathbf{r}_0$ and $\mathbf{p}_0$ for brevity. Likewise, continuity of the particle translational flux across the cell edges implies
\begin{align}
	&\nabla_{r} \psi(\mathbf{R_{\alpha\beta}},\mathbf{r},\mathbf{p},t)= \nabla_{r}\psi(\mathbf{R}_{\alpha+1\beta},\mathbf{r}-L_x\mathbf{e}_x,\mathbf{p},t)  \quad \text{on} \quad \mathbf{r} \in \partial C_{\alpha^+ \beta}, \label{eq:BC_dpsi_edgesip}\\
	&\nabla_{r}\psi(\mathbf{R_{\alpha\beta}},\mathbf{r},\mathbf{p},t)= \nabla_{r}\psi(\mathbf{R}_{\alpha\beta+1},\mathbf{r}-L_y\mathbf{e}_y,\mathbf{p},t)  \quad \text{on} \quad \mathbf{r} \in \partial C_{\alpha\beta^+}. \label{eq:BC_dpsi_edgesjp}
\end{align}
These continuity relations will be useful in the following discussion when deriving boundary conditions for the probability moments.

\subsection{Macrotransport model}

The Fokker-Planck description forms the basis for the calculation of the asymptotic mean transport velocity $\overline{\mathbf{U}}$ and dispersivity dyadic $\overline{\mathbf{D}}$ as we now proceed to explain. The method described here is an extension of  Brenner's generalized Taylor dispersion theory \cite{brenner1980dispersion} to the case of polar active particles with a unit director $\mathbf{p}$ undergoing rotational diffusion. 

\subsubsection{Local moments}

Following generalized Taylor dispersion theory \cite{brenner1980dispersion}, we seek the mean asymptotic transport properties in terms of the $k$th-order polyadic local moments of $\psi$ defined as 
\begin{equation}
\boldsymbol{\psi}_k(\mathbf{r},\mathbf{p},t|\mathbf{r}_0,\mathbf{p}_0)=\sum_{\alpha,\beta=-\infty}^{\infty}  \underbrace{\mathbf{R}_{\alpha\beta}\cdot\cdot\cdot\mathbf{R}_{\alpha\beta}}_{k\text{ times}}\, \psi\,. \label{eq:local_moment} 
\end{equation}
Taking moments of the Fokker-Planck equation \eqref{eq:FP_local} provides conservation equations for the $\boldsymbol{\psi}_k$'s:
\begin{equation}
\frac{\partial \boldsymbol{\psi}_k}{\partial t}+\nabla_r \cdot \mathbf{J}_k+\nabla_p \cdot \mathbf{j}_k=\delta_{k0}\delta(\mathbf{r}-\mathbf{r}_0)\delta(\mathbf{p}-\mathbf{p}_0) \delta(t), \label{eq:psik} 
\end{equation}
where $\mathbf{J}_k$ and $\mathbf{j}_k$ are the local moments of the translational and orientational fluxes, respectively, and are simply given by:
\begin{align}
	\mathbf{J}_k &=\left[\mathrm{Pe}_s \mathbf{p}+ \mathrm{Pe}_f\mathbf{u}(\mathbf{r})\right] \boldsymbol{\psi}_k-\kappa^2 \nabla_r \boldsymbol{\psi}_k,  \label{eq:xk_flux}\\
	\mathbf{j}_k &=\frac{1}{2}\mathrm{Pe}_f \left[ \boldsymbol{\Omega}(\mathbf{r})\times \mathbf{p}\right] \boldsymbol{\psi}_k -\nabla_p \boldsymbol{\psi}_k. \label{eq:pk_flux}
\end{align}
The pre-initial condition on $\psi$ implies that $\boldsymbol{\psi}_k=\mathbf{0}$ for $t<0$. Eq.~(\ref{eq:no-flux}) results in a no-flux condition for the local moments at the obstacle walls: $\mathbf{n}_w \cdot \mathbf{J}_k=0$ on $\partial C_w$. As we will show later, knowledge of the first three local moments $(k = 0,1,2)$ is sufficient to determine the asymptotic transport velocity and dispersion dyadic. 

The continuity conditions obtained in Eqs.\ \eqref{eq:BC_psi_edgesip}--\eqref{eq:BC_psi_edgesjp} can be used to derive boundary conditions for the moments. Upon taking the local moments of \eqref{eq:BC_psi_edgesip}--\eqref{eq:BC_psi_edgesjp} and applying translational invariance of Eq.~\eqref{eq:local_moment} with respect to indices $(\alpha,\beta)$, we obtain:
\begin{align}
\psi_0(\mathbf{r},\mathbf{p},t)&=\psi_0(\mathbf{r}-L_x \mathbf{e}_x,\mathbf{p},t) \quad \text{on} \quad \mathbf{r} \in \partial C_{\alpha^+\beta}. \label{eq:no-flux_psi0k_1} \\
\psi_0(\mathbf{r},\mathbf{p},t)&=\psi_0(\mathbf{r}-L_y \mathbf{e}_y,\mathbf{p},t) \quad \text{on} \quad \mathbf{r} \in \partial C_{\alpha\beta^+}. \label{eq:no-flux_psi0k_2}\vspace{-0.2cm}
\end{align}
For the first local moment we find:
\begin{align}
&\boldsymbol{\psi}_1(\mathbf{r})-\boldsymbol{\psi}_1(\mathbf{r}-L_x \mathbf{e}_x)=(\mathbf{r}-L_x \mathbf{e}_x) \psi_0(\mathbf{r}-L_x \mathbf{e}_x)  -\mathbf{r} \psi_0(\mathbf{r}) \quad \text{on} \,\, \mathbf{r} \in \partial C_{\alpha^+\beta}, \hspace{-0.2cm}
\label{eq:no-flux_psi1k_2} \\
&\boldsymbol{\psi}_1(\mathbf{r})-\boldsymbol{\psi}_1(\mathbf{r}-L_y \mathbf{e}_y)=(\mathbf{r}-L_y \mathbf{e}_y) \psi_0(\mathbf{r}-L_y \mathbf{e}_y)  -\mathbf{r} \psi_0(\mathbf{r}) \quad \text{on} \,\, \mathbf{r} \in \partial C_{\alpha\beta^+},  \label{eq:no-flux_psi1k_3} 
\end{align}
and similarly for the second moment,
\begin{align}
&\boldsymbol{\psi}_2(\mathbf{r})-\boldsymbol{\psi}_2(\mathbf{r}-L_x\mathbf{e}_x)=\frac{\boldsymbol{\psi}_1(\mathbf{r})\boldsymbol{\psi}_1(\mathbf{r})}{{\psi_0}(\mathbf{r})}-\frac{\boldsymbol{\psi}_1(\mathbf{r}-L_x\mathbf{e}_x)\boldsymbol{\psi}_1(\mathbf{r}-L_x\mathbf{e}_x)}{{\psi_0}(\mathbf{r}-L_x\mathbf{e}_x)} \quad \text{on} \,\, \mathbf{r} \in \partial C_{\alpha^+ \beta}, 
\label{eq:no-flux_psi2k_2} \\
&\boldsymbol{\psi}_2(\mathbf{r})-\boldsymbol{\psi}_2(\mathbf{r}-L_y\mathbf{e}_y)=\frac{\boldsymbol{\psi}_1(\mathbf{r})\boldsymbol{\psi}_1(\mathbf{r})}{{\psi_0}(\mathbf{r})}-\frac{\boldsymbol{\psi}_1(\mathbf{r}-L_y\mathbf{e}_y)\boldsymbol{\psi}_1(\mathbf{r}-L_y\mathbf{e}_y)}{{\psi_0}(\mathbf{r}-L_y\mathbf{e}_y)} \quad \text{on} \,\, \mathbf{r} \in \partial C_{\alpha \beta^+}, 
\label{eq:no-flux_psi2k_3} \vspace{-0.2cm}
\end{align}
where we have omitted the dependence on $\mathbf{p}$ and $t$ for brevity. To further simplify the notations, we define the jump $\jump{\mathbf{f}}$ of a variable $\mathbf{f}(\mathbf{r},\mathbf{p},t)$ as
\begin{equation}
\jump{\mathbf{f}}=\left\{
\begin{array}{ll}
\mathbf{f}(\mathbf{r},\mathbf{p},t)-\mathbf{f}(\mathbf{r}-L_x\mathbf{e}_x,\mathbf{p},t) \quad \text{for} \quad \mathbf{r} \in \partial C_{r}, \\
\mathbf{f}(\mathbf{r},\mathbf{p},t)-\mathbf{f}(\mathbf{r}-L_y\mathbf{e}_y,\mathbf{p},t) \quad \text{for} \quad \mathbf{r} \in \partial C_{t},
\end{array}
\right.  \label{eq:jump_operator}
\end{equation}
where $\partial C_r$ and $\partial C_t$ are defined in Fig.~\ref{fig:geo}($a$).
With this notation, the boundary conditions of Eqs.\ (\ref{eq:no-flux_psi0k_1})--(\ref{eq:no-flux_psi2k_3}) take on the simple form
\begin{align}
	&\jump{\psi_0}=0, \label{eq:psi0_BC_jump1}\\
	&\jump{\boldsymbol{\psi}_1}=-\jump{\mathbf{r}\psi_0}, \label{eq:psi1_BC_jump1}\\
	&\jump{\boldsymbol{\psi}_2}=\jump{\boldsymbol{\psi}_1\boldsymbol{\psi}_1/\psi_0}.  \label{eq:psi2_BC_jump1} 
\end{align}
Following a similar procedure, we can use Eqs.\ (\ref{eq:BC_dpsi_edgesip})--(\ref{eq:BC_dpsi_edgesjp}) to derive boundary conditions for the fluxes, which read
\begin{align}
	&\jump{\nabla_r\psi_0}=\mathbf{0}, \label{eq:psi0_BC_jump2}\\
	&\jump{\nabla_r \boldsymbol{\psi}_1}=-\jump{\nabla_r\left(\mathbf{r}\psi_0\right)}, \label{eq:psi1_BC_jump2}\\
	&\jump{\nabla_r \boldsymbol{\psi}_2}=\jump{\nabla_r \left(\boldsymbol{\psi}_1\boldsymbol{\psi}_1/\psi_0\right)}.  \label{eq:psi2_BC_jump2} 
\end{align}
These conditions are identical to those obtained by Brenner \cite{brenner1980dispersion} for passive Brownian particles. In the following sections we will take the asymptotic limits of the local moments. With that in mind, it will be useful to rewrite the first-order jump conditions (\ref{eq:psi1_BC_jump1}) and (\ref{eq:psi1_BC_jump2}) in the form:
\begin{equation}
\jump{\boldsymbol{\psi}_1}=-\psi_0\,\jump{\mathbf{r}}, \qquad \jump{\nabla_r \boldsymbol{\psi}_1}=-\nabla_r \psi_0 \,\jump{\mathbf{r}},  \label{eq:BCs_psi1} 
\end{equation}
where we have used the zeroth-order conditions (\ref{eq:psi0_BC_jump1}) and (\ref{eq:psi0_BC_jump2}).

\subsubsection{Global moments}

We now proceed to define the $k$th-order dyadic global moment $\mathbf{M}_k$ as the spatial and orientational average of the corresponding local moment $\boldsymbol{\psi}_k$:\vspace{-0.1cm}
\begin{equation}
\mathbf{M}_k(t|\mathbf{r}_0,\mathbf{p}_0)=\int_{S_{t} } \int_{\Omega} \boldsymbol{\psi}_k(\mathbf{r},\mathbf{p},t|\mathbf{r}_0,\mathbf{p}_0) \,d\mathbf{p}\, d^2\mathbf{r},  \label{eq:Mk} \quad k=0,1,2... \vspace{-0.1cm} 
\end{equation}
This definition is analogous to that of Brenner \cite{brenner1980dispersion}, with the main difference that $\mathbf{p}$ serves as an additional local variable for orientable particles \cite{brenner1979}. The zeroth-order global moment derives immediately from the normalization condition of Eq. (\ref{eq:norma_cond}):\vspace{-0.1cm}
\begin{equation}
M_0=\left\{
\begin{array}{ll}
1 \quad \text{for} \quad t \ge 0,\\
0 \quad \text{for} \quad t < 0.
\end{array}
\right. \label{eq:Mk0} 
\end{equation}
A differential equation for $\mathbf{M}_k(t)$ can be obtained by taking the temporal derivative of Eq.~(\ref{eq:Mk}) and combining it with Eq.~(\ref{eq:psik}):
\begin{equation}
\frac{d \mathbf{M}_k}{dt}=\int_{S_{t} } \int_{\Omega} \left[-\nabla_r \cdot \mathbf{J}_k-\nabla_p \cdot \mathbf{j}_k+\delta_{k0}\delta(\mathbf{r}-\mathbf{r}_0)\delta(\mathbf{p}-\mathbf{p}_0) \delta(t)\right] d\mathbf{p} \, d^2\mathbf{r}.
\end{equation}  
The integral of the orientational flux divergence vanishes. Applying the divergence theorem then gives:\vspace{-0.1cm}
\begin{equation}
\frac{d \mathbf{M}_k}{dt}= -   \int_{\Omega}\oint_{\partial C } \mathbf{J}_k\cdot \mathbf{n} \,d\ell\,  d\mathbf{p}  +\delta_{k0}\delta(t), \label{eq:Governing_Mk}\vspace{-0.1cm}
\end{equation}
where we have used the no-flux condition on the surface of the obstacles. In Eq.~(\ref{eq:Governing_Mk}), the closed curve integral with line element $d\ell$ represents the sum of the integrals over all the unit cell edges shown in Fig.~\ref{fig:geo}($a$):\vspace{-0.1cm}
\begin{equation}
\oint_{\partial C }= \int_{\partial C_{r}}+\int_{\partial C_{t}}+\int_{\partial C_{l}}+\int_{\partial C_{b}} .\vspace{-0.1cm}
\end{equation}  
By making use of the jump operator defined in Eq.~(\ref{eq:jump_operator}), Eq. (\ref{eq:Governing_Mk}) is conveniently rewritten as
\begin{equation}
\frac{d\mathbf{M}_k}{dt}=-\int_{\Omega} \int_{\partial C_{r}+\partial C_{t}}\jump{\mathbf{J}_k} \cdot\mathbf{n}\, d\ell \, d\mathbf{p}+\delta_{k0} \delta(t). \label{eq:Governing_Mk2}
\end{equation}

\subsubsection{Asymptotic analysis at long times\label{sec:asymptotic}}

Over time, the active particle is transported across the array and samples many lattice cells. In the long-time asymptotic limit, we seek to describe its statistics in terms of its mean transport velocity $\overline{\mathbf{U}}$ and mean dispersivity dyadic $\overline{\mathbf{D}}$. As we now explain, these two quantities are easily obtained in terms of the probability moments introduced previously.

\paragraph{{Mean velocity:}}To determine the mean velocity $\overline{\mathbf{U}}$, we resort to a Lagrangian description and calculate the ensemble-averaged swimmer displacement at time $t$ as
\begin{equation}
\langle{\Delta \mathbf{R}}(t)\rangle=\langle{\mathbf{R}(t)-\mathbf{R}(0)}\rangle=\sum_{\alpha,\beta=-\infty}^{\infty} \int_{S_{\alpha\beta}} \int_{\Omega} \mathbf{R_{\alpha\beta}} \, \psi(\mathbf{R_{\alpha\beta}},\mathbf{r},\mathbf{p},t| \mathbf{r}_0,\mathbf{p}_0) \, d\mathbf{p}\, d^2\mathbf{r}. \label{eq:MeanDisplacement}
\end{equation}
From the definition of global moments it is easy to recognize that the right-hand side in Eq.~(\ref{eq:MeanDisplacement}) is nothing but $\mathbf{M}_1$. Therefore the asymptotic velocity can be expressed as
\begin{equation}
\overline{\mathbf{U}}=\lim_{t\rightarrow\infty}\frac{d}{dt}\langle{\Delta \mathbf{R}}(t)\rangle=\lim_{t\rightarrow\infty}\frac{d\mathbf{M}_1}{dt}=-\lim_{t\rightarrow\infty}\int_{\Omega} \int_{\partial C_{r}+\partial C_{t}}\jump{\mathbf{J}_1} \cdot \mathbf{n}\,d\ell\, d\mathbf{p}, \label{eq:velexpression}
\end{equation}
where the last relation results from Eq.~(\ref{eq:Governing_Mk2}). Using Eq.~\eqref{eq:BCs_psi1} and the definition of the moment flux in Eq.~\eqref{eq:xk_flux}, it is straightforward to show that $\jump{\mathbf{J}_1}=-\mathbf{J}_0 \jump{\mathbf{r}}$, and therefore
\begin{equation}
\mathbf{\overline{U}}= \int_{\Omega}  \int_{\partial C_{r}+\partial C_{t}} \jump{\mathbf{r}}\mathbf{J}_0^{\infty} \cdot  \mathbf{n}\,d\ell \, d\mathbf{p}.  \label{eq:MeanVelocity}
\end{equation}
where $\mathbf{J}_0^\infty (\mathbf{r},\mathbf{p})\equiv \lim_{t\rightarrow \infty} \mathbf{J}_0(\mathbf{r},\mathbf{p},t)$, with an asymptotic approach that can be shown to be exponential in time  \cite{brenner1980dispersion}.
The mean asymptotic velocity thus solely depends on the steady-state zeroth-order local moment $\psi_0^{\infty}(\mathbf{r},\mathbf{p})$, which describes the long-time probability of finding the particle at local position $\mathbf{r}$ with orientation $\mathbf{p}$ irrespective of which unit cell it is traversing.\ It satisfies the simplified steady Fokker-Planck equation:
\begin{equation}
\smash{\nabla_r\cdot\mathbf{J}_{0}^\infty+\nabla_p\cdot\mathbf{j}_{0}^\infty=0}, \label{eq:psi0_longt}
\end{equation}
with fluxes
\begin{align}
\mathbf{J}_{0}^\infty&=\left[\mathrm{Pe}_s \mathbf{p}+ \mathrm{Pe}_f \mathbf{u}(\mathbf{r})\right]\psi_0^\infty-\kappa^2\nabla_r {\psi_0^{\infty}},   \label{eq:x0_flux_longt}\\
\mathbf{j}_0^{\infty}&= \frac{1}{2}\mathrm{Pe}_f\left[\boldsymbol{\Omega}(\mathbf{r})\times \mathbf{p}\right]\psi_0^\infty-(\mathbf{I}-\mathbf{pp})\cdot\nabla_p {\psi_0^{\infty}},  \label{eq:p0_flux_longt}  \vspace{-0.15cm}
\end{align}
subject to normalization condition $\int_{S }\int_{\Omega}\psi_0^{\infty} d\mathbf{p}\, d^2\mathbf{r}=1$, to the no-flux condition $\mathbf{n}_w\cdot \mathbf{J}_0^{\infty}=0$ on the surface of the pillar, and to jump conditions $\jump{\psi_0^{\infty}}=0$ and $\jump {\nabla_r \psi_0^{\infty}}=\mathbf{0}$ on the unit cell edges $\partial C_{r}$ and $\partial C_{t}$.\ In the results shown in Sec.~\ref{sec:results}, we solve for $\psi_0^{\infty}$ numerically using a finite-volume method. Once $\psi_0^\infty$ has been obtained, we calculate $\overline{\mathbf{U}}$ by simple quadrature according to Eq.~\eqref{eq:MeanVelocity}, where the long-time translational flux is given in Eq.~(\ref{eq:x0_flux_longt}).

\paragraph{{Mean dispersivity:}} To determine the dispersivity tensor $\overline{\mathbf{D}}$, we now consider the ensemble-averaged mean-square displacement of a swimmer:
\begin{equation}
\langle{\left(\Delta \mathbf{R}(t)-\langle{\Delta \mathbf{R}}(t)\rangle\right)^2}\rangle=\sum_{\alpha,\beta=-\infty}^{\infty} \int_{S_{\alpha\beta}} \int_{\Omega} \left(\mathbf{X_{\alpha\beta}}- \langle\Delta \mathbf{R}(t)\rangle\right)^2\psi(\mathbf{X_{\alpha\beta}},\mathbf{r},\mathbf{p},t| \mathbf{r}_0,\mathbf{p}_0) \, d\mathbf{p} \,d^2\mathbf{r}. \label{eq:MeanSquareDisplacement}
\end{equation}
Expanding the right-hand side and using the definition of the global moments yields
\begin{align}
\begin{split}
\langle{\left(\Delta \mathbf{R}(t)-\langle{\Delta \mathbf{R}}(t)\rangle\right)^2}\rangle&=\mathbf{M}_2(t)+\langle\Delta \mathbf{R}(t)\rangle^2 M_0-\mathbf{M}_1(t)\langle{\Delta \mathbf{R}}(t)\rangle-\langle{\Delta \mathbf{R}}(t)\rangle \mathbf{M}_1(t)\\&=\mathbf{M}_2(t)-\mathbf{M}_1(t)\mathbf{M}_1(t),\label{eq:MeanSquareDisplacement2}
\end{split}
\end{align}
where we have used $\langle\mathbf{R}(t)\rangle=\mathbf{M}_1(t)$ and $M_0=1$.\ The asymptotic dispersivity tensor can then be obtained as
\begin{equation}
\overline{\mathbf{D}}=\lim_{t\to \infty}\frac{1}{2}\frac{d}{dt}\langle{\left(\Delta \mathbf{R}(t)-\langle{\Delta \mathbf{R}}(t)\rangle\right)^2}\rangle= \lim_{t\to \infty} \frac{1}{2}\frac{d}{dt}\left(\mathbf{M}_2-\mathbf{M}_1 \mathbf{M}_1\right).  \label{eq:Dispersivity}
\end{equation}
To make further progress, we first note that integrating Eq.~(\ref{eq:velexpression}) with respect to time gives
\begin{equation}
\mathbf{M}_1(t)=\mathbf{\overline{U}} t +\overline{\mathbf{B}}+\mathcal{O}(\mathrm{e}^{-t}), \label{eq:M1}
\end{equation}
where $\overline{\mathbf{B}}$ is an unknown constant vector. Let us recall the definition of $\mathbf{M}_1(t)$:
\begin{equation}
\mathbf{M}_1(t)=\int_{S_{t}}\int_{\Omega} \boldsymbol{\psi}_1(\mathbf{r},\mathbf{p},t)\,d\mathbf{p}\, d^2\mathbf{r}. \label{eq:M1_relation}
\end{equation}
Comparison of Eqs. (\ref{eq:M1}) and (\ref{eq:M1_relation}) suggests seeking the asymptotic solution for $\boldsymbol{\psi}_1$ in the form
\begin{equation}
\boldsymbol{\psi}_1(\mathbf{r},\mathbf{p},t)=\psi_0^{\infty}(\mathbf{r},\mathbf{p})\left[\mathbf{\overline{U}} t +\mathbf{B}(\mathbf{r},\mathbf{p})\right]+\mathcal{O}(\mathrm{e}^{-t}), \label{eq:psi1_asymptotic}
\end{equation}
where $\mathbf{B}(\mathbf{r},\mathbf{p})$ is a vector field to be determined such that
\begin{equation}
\overline{\mathbf{B}}=\int_{S_{t}}\int_{\Omega} \psi_0^{\infty}\mathbf{B}(\mathbf{r},\mathbf{p})\,d\mathbf{p} \,d^2\mathbf{r}. \label{eq:Bbar}
\end{equation}
The $\mathcal{O}(\mathrm{e}^{-t})$ term in Eqs.~\eqref{eq:M1} and \eqref{eq:psi1_asymptotic} is a consequence of the exponential convergence of the zeroth-order local moment with time \cite{brenner1980dispersion}. Using Eq.~(\ref{eq:M1_relation}), we can rewrite Eq.~(\ref{eq:Dispersivity}) as
\begin{equation}
\overline{\mathbf{D}}=\lim_{t\to \infty} \frac{1}{2}\frac{d \mathbf{M}_2}{dt}-\overline{\mathbf{U}} \,\overline{\mathbf{U}} t-\frac{1}{2}\left(\overline{\mathbf{U}} \,\overline{\mathbf{B}}+\overline{\mathbf{B}} \,\overline{\mathbf{U}}\right). \label{eq:Dispersivity2}
\end{equation}
The first term on the right-hand side is estimated using the evolution equation \eqref{eq:Governing_Mk2} for the global moments:
\begin{equation}
\lim_{t\to \infty}\frac{d \mathbf{M}_2}{dt}=-\lim_{t\to \infty} \int_{\Omega}\int_{\partial C_{r}+\partial C_{t}} \jump{\mathbf{J}_2}\cdot \mathbf{n}\,d\ell\, d\mathbf{p}.  \label{eq:dM2dt}
\end{equation}
The jump in $\mathbf{J}_2$ can be obtained by combining Eq.~(\ref{eq:xk_flux}) with the boundary conditions (\ref{eq:psi2_BC_jump1})--(\ref{eq:psi2_BC_jump2}) along with Eq.~(\ref{eq:psi1_asymptotic}), yielding
\begin{equation}
\lim_{t\to \infty} \jump{\mathbf{J}_2}=-\mathbf{J}_0^{\infty}\left(\overline{\mathbf{U}}\jump{\mathbf{r}} t+\jump{\mathbf{r}}\overline{\mathbf{U}}t-\jump{\mathbf{B}\mathbf{B}} \right) -\kappa^2 \psi_0^{\infty}\jump{\nabla_r \left(\mathbf{B}\mathbf{B}\right)}. \label{eq:jumpx2_flux_simpl}
\end{equation}
Upon inserting Eqs.~(\ref{eq:jumpx2_flux_simpl}) and (\ref{eq:dM2dt}) into Eq.~(\ref{eq:Dispersivity2}), we obtain a new expression for the dispersivity,
\begin{equation}
\overline{\mathbf{D}}=\frac{1}{2}\int_{\Omega}\int_{\partial C_{r}+\partial C_{t}} \left[\kappa^2 \psi_0^{\infty}\jump{\nabla_r \left(\mathbf{B}\mathbf{B}\right)}-\mathbf{J}_0^{\infty}\jump{\mathbf{B}\mathbf{B}} \right]\cdot \mathbf{n}\,d\ell\, d\mathbf{p} -\frac{1}{2}\left(\overline{\mathbf{U}} \,\overline{\mathbf{B}}+\overline{\mathbf{B}} \,\overline{\mathbf{U}}\right), \label{eq:Dispersivity3}
\end{equation}
which depends on the known zeroth-order moment $\psi_0^\infty$ as well as on the yet unknown fluctuation field $\mathbf{B}(\mathbf{r},\mathbf{p})$. An equation for $\mathbf{B}$ can be obtained by plugging Eq.~\eqref{eq:psi1_asymptotic} into the governing equation for $\boldsymbol{\psi}_1$ and making use of the jump conditions. The mathematical manipulations are similar to those of Brenner \cite{brenner1980dispersion} and yield the equation
\begin{equation}
-\nabla_r \cdot \left[\mathbf{J}_0^{\infty}  \mathbf{B}-\kappa^2 \psi_0 \nabla_r \mathbf{B}\right] -\nabla_p \cdot \left[\mathbf{j}_0^{\infty} \mathbf{B}- \psi_0 \nabla_p \mathbf{B}\right]=\psi_0^{\infty} \overline{\mathbf{U}}, \label{eq:Bfield}
\end{equation}
subject to boundary conditions $\jump{\mathbf{B}}=-\jump{\mathbf{r}}$ and $\jump{\nabla_r \mathbf{B}}=\mathbf{0}$ on $\partial C_r,\partial C_t$, and $\mathbf{n}_w\cdot \nabla_r\mathbf{B}=\mathbf{0}$ on $\partial C_w$. This problem can also be solved numerically using finite volumes, from which one then obtains $\overline{\mathbf{D}}$ using Eq.~(\ref{eq:Dispersivity3}). 

\subsubsection{Summary of the macrotransport model} 

\begin{figure}[t]
	\centering
	\includegraphics[width=0.4\linewidth]{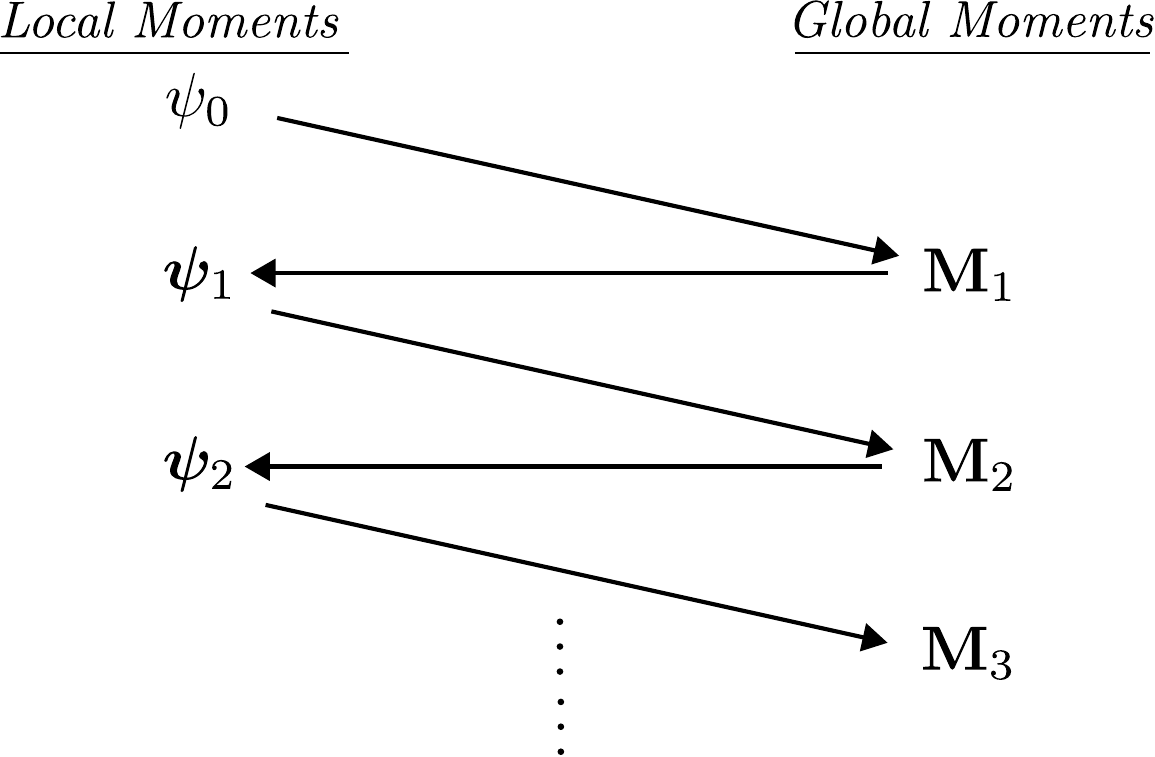}
	\caption{Hierarchical structure for determining the local and global moments of the conditional probability density function $\psi(\mathbf{R},\mathbf{p},t|\mathbf{R}_0,\mathbf{p}_0)$.\ The first and second global moments contain information about asymptotic transport velocity and dispersivity and can be solved for with knowledge of the zeroth and first local moments. Cartoon inspired from \cite{brenner2013macrotransport}.}
	\label{fig:moments}
\end{figure}

We take a moment to summarize and highlight the main steps of the macrotransport model presented above, which extends that of Brenner \cite{brenner1980dispersion} to the case of active particles and relies on the hierarchical structure of the governing equations for the probability moments. As we showed in Eqs.~(\ref{eq:velexpression}) and (\ref{eq:Dispersivity}), the asymptotic transport velocity $\overline{\mathbf{U}}$ depends on the first global moment $\mathbf{M}_1$, while the dispersion dyadic $\overline{\mathbf{D}}$ requires knowledge of the second global moment $\mathbf{M}_2$; higher-order moments of the Lagrangian particle displacement would similarly involve global moments of higher orders, e.g.~$\mathbf{M}_3$. Interestingly, we also saw that the asymptotic form of the first global moment can be obtained with  knowledge of the zeroth-order local moment $\psi_0$, while the second global moment can be solved for using the first-order local moment $\boldsymbol{\psi}_1$.\ This hierarchy of equations is the basis for the method of moments \cite{brenner1980dispersion,edwards1991dispersion}, and we borrow a cartoon from Brenner \cite{brenner2013macrotransport} to illustrate this inter-relation in Fig.~\ref{fig:moments}.  

In order to solve for the mean dispersivity, we need knowledge of the fluctuation field $\mathbf{B}(\mathbf{r},\mathbf{p})$, which obeys advection-diffusion equation \eqref{eq:Bfield} as we derived in the previous section.\ It should be apparent from the boundary conditions that the problem does not have a unique solution and that $\mathbf{B}$ can only be determined up to an arbitrary constant.\ This has to do with the fact that the absolute value of $\mathbf{B}$ does not affect the dispersion tensor, which instead depends on its spatial and orientational gradients. In this sense, $\mathbf{B}$ is often interpreted as a `dispersion potential' \cite{brenner1980dispersion} whose gradients drive the spreading of a cloud of particles. When solving for it numerically using finite volumes, we make the problem solvable by setting an arbitrary gauge condition on the mean value of the field: \vspace{-0.2cm}
\begin{equation}
\int_{S_t}\int_\Omega\mathbf{B}(\mathbf{r},\mathbf{p}) \,d \mathbf{p}\, d^2 \mathbf{r} = \mathbf{0}, \label{eq:zeromean} \vspace{-0.1cm}
\end{equation} 
and this choice does not affect the result for $\overline{\mathbf{D}}$.

\section{Results and Discussion\label{sec:results}}

We now present theoretical results and compare them to discrete Brownian dynamics (BD) simulations based on the Langevin equations (\ref{eq:xdot})--(\ref{eq:pdot}); see Appendix~\ref{app:BD} for details of the algorithm. We first discuss the phenomenology of spreading and convergence to the asymptotic regime, then analyze the effects of geometric and dynamic parameters on long-time microswimmer transport.

\subsection{Darcy-scale transport and dispersion of a cloud}

\begin{figure}[t]
	\centering
	\includegraphics[width=\linewidth]{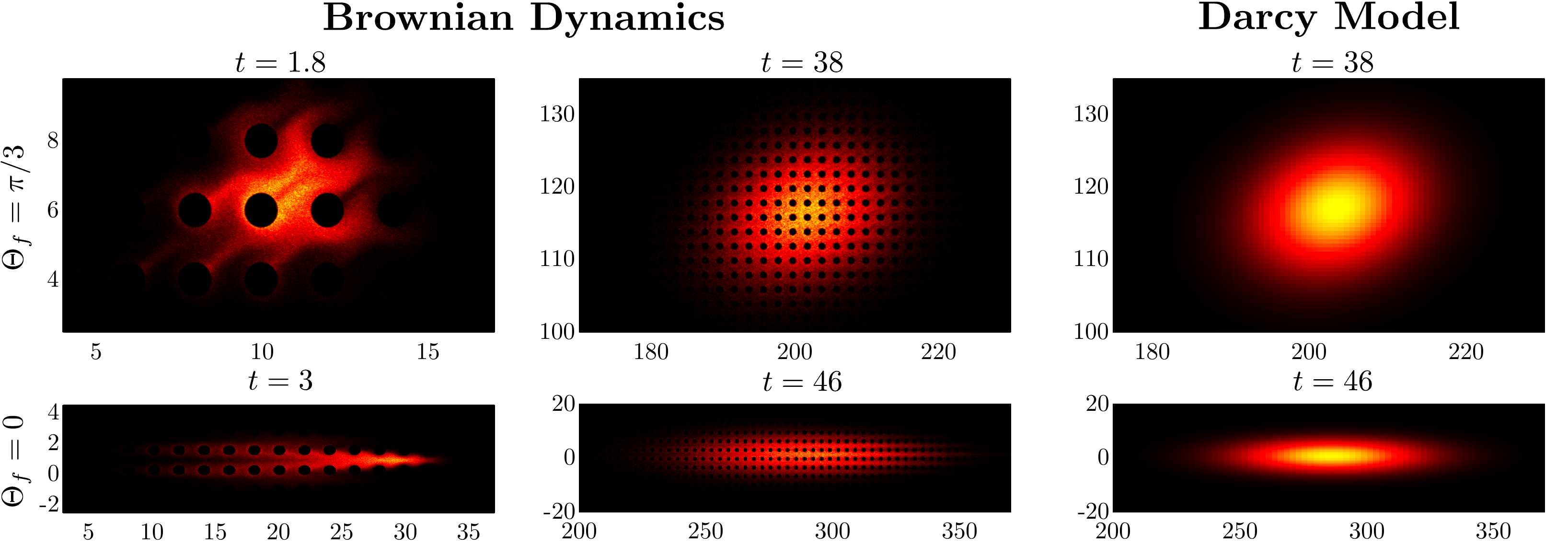}
	\caption{Transport and spreading of a cloud of active particles around circular pillars in Brownian dynamics simulations at two different times (left two columns) compared to the Darcy-scale theoretical prediction of Eq.~(\ref{eq:gaussian}) (third column).\ The two rows correspond to incoming flow angles $\Theta_f=\pi/3$ and~$0$.\ See Supplementary Material for movies showing the temporal evolution.\ Parameter values: $\epsilon_p=0.804$, $\mathrm{Pe}_s=1$, $\kappa^2=0.1$, and $\mathrm{Pe}_f=5$. } \label{fig:Eulerian} 
\end{figure}

The macrotransport model predicts that, on long length and time scales, a dilute cloud of active particles will be transported with mean velocity $\overline{\mathbf{U}}$ and will spread with dispersivity $\overline{\mathbf{D}}$ as provided by Eqs.~(\ref{eq:MeanVelocity}) and (\ref{eq:Dispersivity}).\ This suggests seeking a coarse-grained Eulerian interpretation of $\overline{\mathbf{U}}$ and $\overline{\mathbf{D}}$ at the Darcy scale, where the fluid and solid obstacles that comprise the porous material become indistinguishable and distances shorter than the characteristic size of the unit cell are irrelevant.\ To this end, following Brenner \cite{brenner1980dispersion} we introduce the discrete conditional probability density as:
\begin{equation}
\Psi(\mathbf{R}_{\alpha \beta},t|\mathbf{r}_0,\mathbf{p}_0)= \frac{1}{S_t}\int_{\Omega}\int_{S_{\alpha \beta}} \psi(\mathbf{R}_{\alpha \beta},\mathbf{r},\mathbf{p},t|\mathbf{r}_0,\mathbf{p}_0)\, d^2 \mathbf{r}\, d\mathbf{p},\label{eq:psibar}
\end{equation}
which represents the probability for a particle to be located in the cell labeled by the integers ($\alpha, \beta$) at time $t$, and whose value is assigned to the centroid of the unit cell. \ On large length scales, we can assimilate $\mathbf{R}_{\alpha \beta}$ to a continuous variable $\mathbf{X}$ and define a corresponding macro-scale gradient operator $\nabla_{X}$. The Darcy-scale probability density $\Psi(\mathbf{X},t)$, the continuous counterpart of ${\psi}(\mathbf{R_{\alpha \beta}},t|\mathbf{r}_0,\mathbf{p}_0)$, is then expected to formally satisfy a 2D obstacle-free advection-diffusion equation:
\begin{equation}
\frac{\partial \Psi}{\partial t}+ \nabla_{{{X}}} \cdot \left(\mathbf{\overline{U}}\Psi -\mathbf{\overline{D}} \cdot \nabla_{{{X}}}{\Psi}  \right)=\delta(\mathbf{{X}}-\mathbf{X}_0)\delta(t)\,. \label{eq:adv_diff} 
\end{equation}
While a rigorous derivation of Eq.~(\ref{eq:adv_diff}) is non-trivial, it can easily be verified that its solution has the same global moments as $\psi$. For an arbitrary incoming flow, the dispersion tensor is non-diagonal but can be expressed as $\overline{\mathbf{D}}=D_1\mathbf{e}_1\mathbf{e}_1+D_2\mathbf{e}_2\mathbf{e}_2$ where $(D_1,D_2)$ are its eigenvalues with corresponding eigenvectors $(\mathbf{e}_1,\mathbf{e}_2)$. With these notations, the solution of Eq.~(\ref{eq:adv_diff}) for the spreading of a concentrated point source can be written as
\begin{equation}
\Psi(\mathbf{X},t) = \frac{1}{4 \pi t \sqrt{D_1D_2}} \exp\left[-\frac{\left(X_1-\overline{U}_1t\right)^2}{4 D_1 t} -\frac{\left(X_2-\overline{U}_2 t\right)^2}{4 D_2 t}\right], \label{eq:gaussian}
\end{equation}
where $(X_1,X_2)$ and $(\overline{U}_1,\overline{U}_2)$ are the components of $\mathbf{X}$ and $\overline{\mathbf{U}}$ in the basis of the eigenvectors. Eq.~(\ref{eq:gaussian}) simply states that the particle cloud will spread as a translating anisotropic Gaussian. 

Results from BD simulations for the spreading of a concentrated point source of active swimmers are compared to the  analytical solution of Eq.~(\ref{eq:gaussian}) in Fig.~\ref{fig:Eulerian} for two flow directions $\Theta_f$ (also see movies in the Supplementary Material).\ On short time scales, the cloud develops a complex shape that is continually distorted by the pillars and exhibits wakes, boundary layers, and other features typical of advective-diffusive transport around obstacles.\ At later times, the cloud shape becomes increasingly smooth and is very well captured by the theoretical prediction.\ This comparison asserts the strength of the Darcy-scale interpretation for describing asymptotic transport, but also highlights key differences.

\begin{figure}[t]
	\centering
	\includegraphics[width=0.9\linewidth]{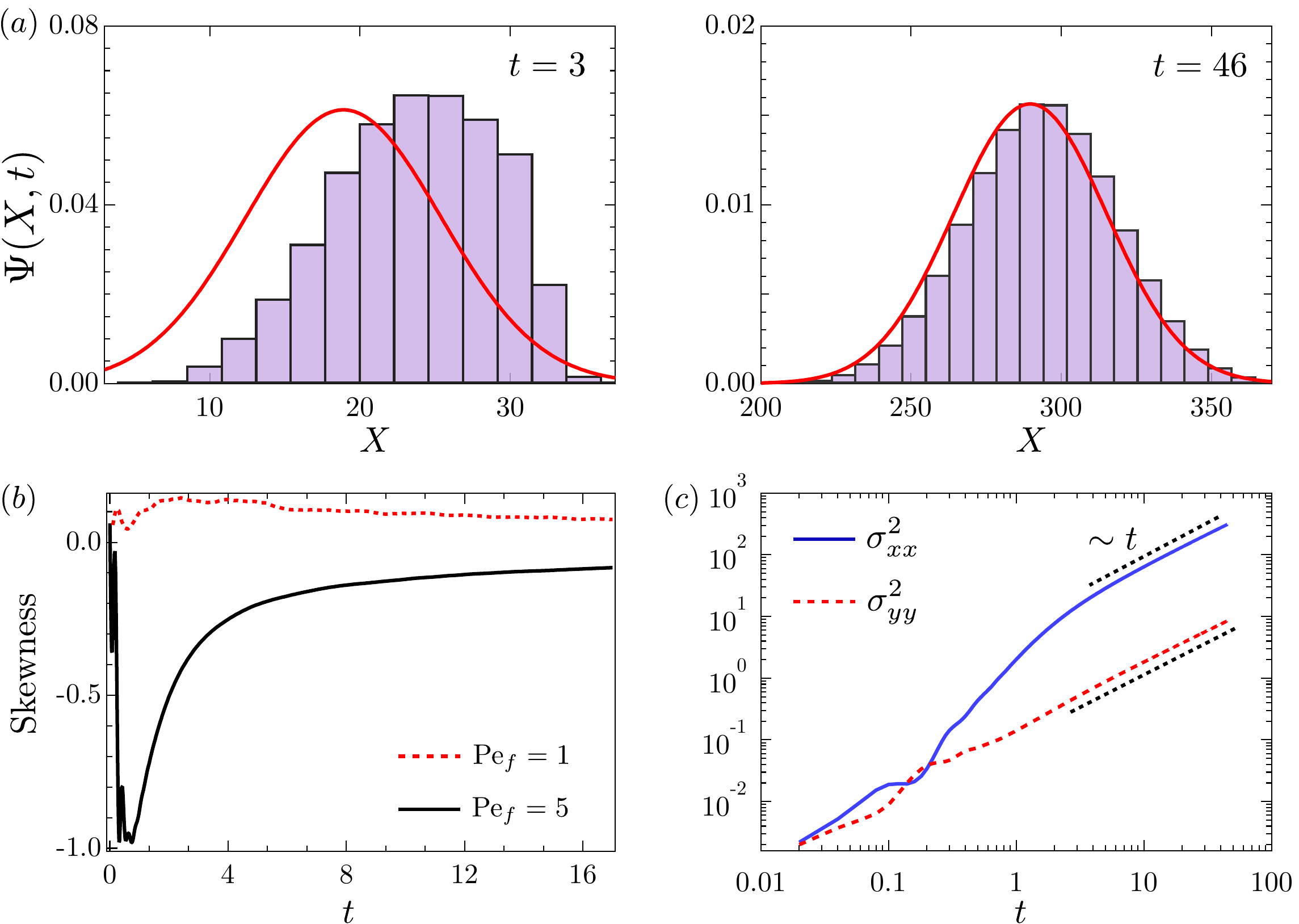}\vspace{-0.3cm}
	\caption{($a$) Probability density function {$\Psi(X,t)$} of the macroscale horizontal position $X$ for the two {instantaneous times} shown in Fig.~\ref{fig:Eulerian} for $\Theta_f=0$, where analytical predictions are compared to BD simulation results.\  (\textit{b}) Evolution of the skewness of particle distributions over time as calculated from Brownian dynamics simulations for two different $\mathrm{Pe}_f$. (\textit{c}) Time evolution of the variances $\sigma_{xx}^2$ and $\sigma_{yy}^2$ of particle positions, showing convergence to the asymptotic diffusive regime.\ In all panels, $\epsilon_p=0.804$, $\mathrm{Pe}_s=1$, $\kappa^2=0.1$, and for {($a$) and ($c$)} $\mathrm{Pe}_f = 5.$ } \label{fig:distributions} 
\end{figure}

Figure~\ref{fig:distributions}(\textit{a}) shows the probability density function along the $X$ direction for the case $\Theta_f = 0$ and provides a more quantitative comparison between Brownian dynamics and the analytical solution. In particular, BD simulations exhibit tailed distributions at pre-asymptotic times, which we characterize by studying the evolution of the skewness in Fig.~\ref{fig:distributions}(\textit{b}). 
A recent experimental study using \textit{E. Coli} in unstructured pillar arrays has reported positively skewed particle distributions \cite{creppy2018motility}.\ Our simulations show more complex trends. In strong flows (large $\mathrm{Pe}_f$), we find that the distributions are always negatively skewed in the direction of the flow with more particles accumulating downstream. However, for fast swimming (large $\mathrm{Pe}_s$) or weak flow (small $\mathrm{Pe}_f$) the distributions are found to be positively skewed. This competition between swimming and flow and its role in transport is elaborated in more detail in the following sections.\ We observe that as time progresses the skewness tends toward zero as the distribution converges towards a Gaussian cloud.\ This convergence is further highlighted by the time evolution of the variance of particle positions in Fig.~\ref{fig:distributions}(\textit{c}), which rapidly reaches the asymptotic diffusive regime of linear growth.

\begin{figure}[t]
	\centering
	\includegraphics[width=0.85\linewidth]{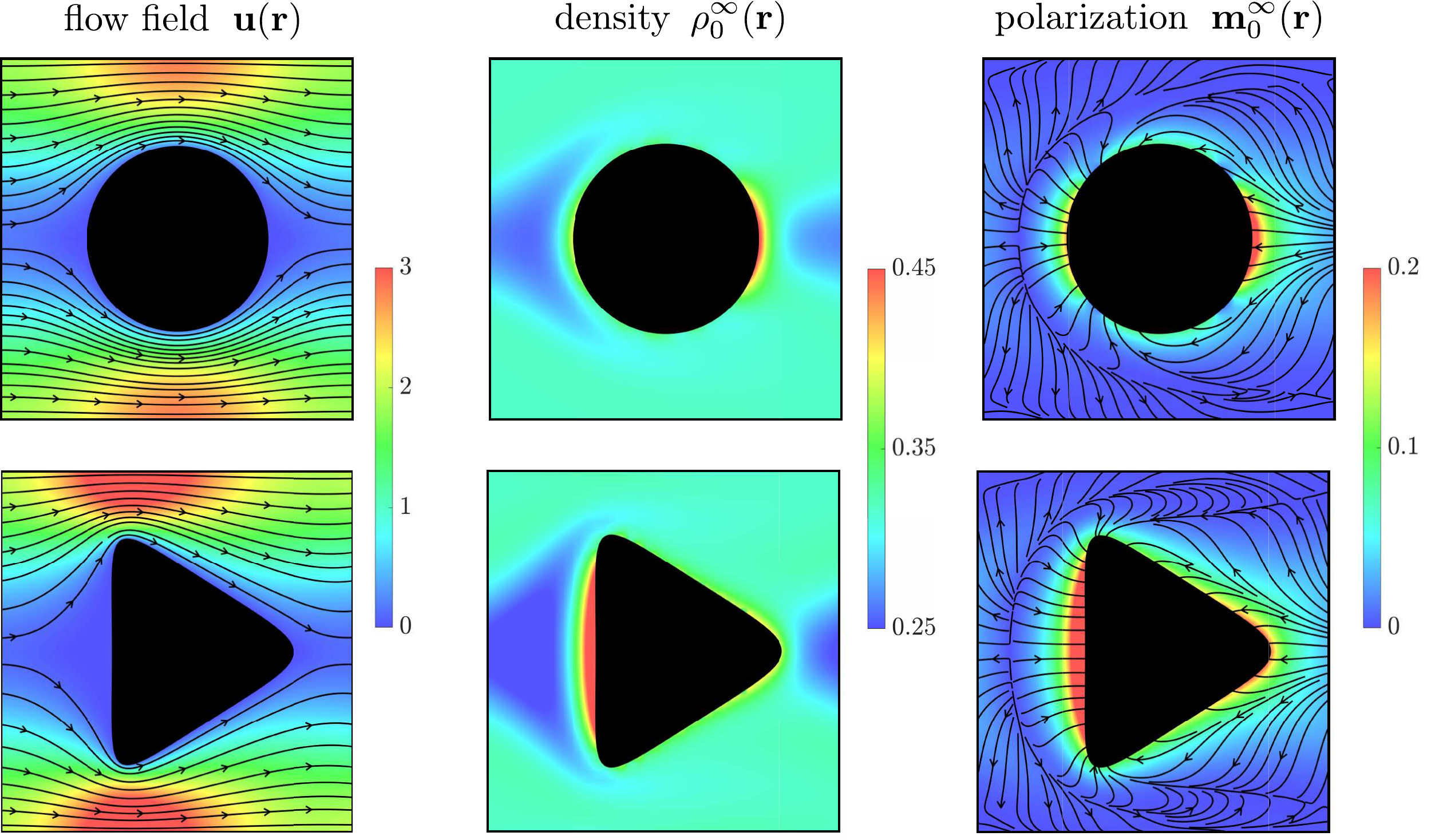}
	\caption{ Asymptotic solutions on the pore scale for two pillar shapes: streamlines and magnitude of the imposed flow $\mathbf{u}(\mathbf{r})$ (\textit{left}), density field $\rho^\infty_0(\mathbf{r})$ (\textit{middle}), and polarization field $\mathbf{m}^\infty_0(\mathbf{r})$ (\textit{right}) as defined in Eq.~(\ref{eq:densitypolarization}).\ In all panels, $\epsilon_p=0.804$, $\mathrm{Pe}_s=1$, $\kappa^2=0.1$, and $\mathrm{Pe}_f=5$.}
	\label{fig:fields}
\end{figure}

\subsection{Asymptotic particle distributions at the pore scale}

Asymptotic transport characteristics can be attributed in part to pore-scale features of the fluid flow and of the steady-state zeroth-order moment $\psi_0^{\infty}$, both of which are illustrated in Fig.~\ref{fig:fields}. We define the density and unnormalized polarization of the particle distribution as the zeroth- and first-order orientational moments of $\psi_0^{\infty}$: 
\begin{align}
\rho^\infty_0(\mathbf{r}) =\int_\Omega \psi_0^{\infty}(\mathbf{r},\mathbf{p})\,d\mathbf{p}, \qquad
\mathbf{m}^{\infty}_0(\mathbf{r})=\int_\Omega \mathbf{p}\,\psi_0^{\infty}(\mathbf{r},\mathbf{p})\,d\mathbf{p}. \label{eq:densitypolarization}
\end{align}
In the absence of flow, the interplay of self-propulsion and translational diffusion near obstacle boundaries creates a net swimmer polarization
against the obstacles \cite{ezhilan2015channel,ezhilan2015b,yan2015boundary}, which vanishes at the cell edges by symmetry.\ This polarization is accompanied by a net increase in density, which is azimuthally symmetric near a circular obstacle and enhanced in regions of high curvature for non-circular pillars \cite{yan2018curved}.\ When a flow is applied, the fore-aft symmetry is broken and particles now preferentially accumulate near stagnation points of the flow, with a maximum density reached in the rear of circular obstacles.\ This downstream accumulation, which is observed in experiments \cite{mino2018coli} and is akin to that occurring past funnel constrictions \cite{altshuler2013flow}, results from the combination of advection by the flow and of shear-rotation and swimming in the fast-flowing high-shear regions above and below the pillars.\ Advection and rotation also induce a density minimum a short distance upstream of the obstacle, which coincides with a stagnation point in the polarization field.\ As discussed in the following sections, all these features play a central role in determining the effective spreading behavior. 

\subsection{Effects of background flow and swimming activity}

\begin{figure}[t]
	\centering
	\includegraphics[width=0.9\linewidth]{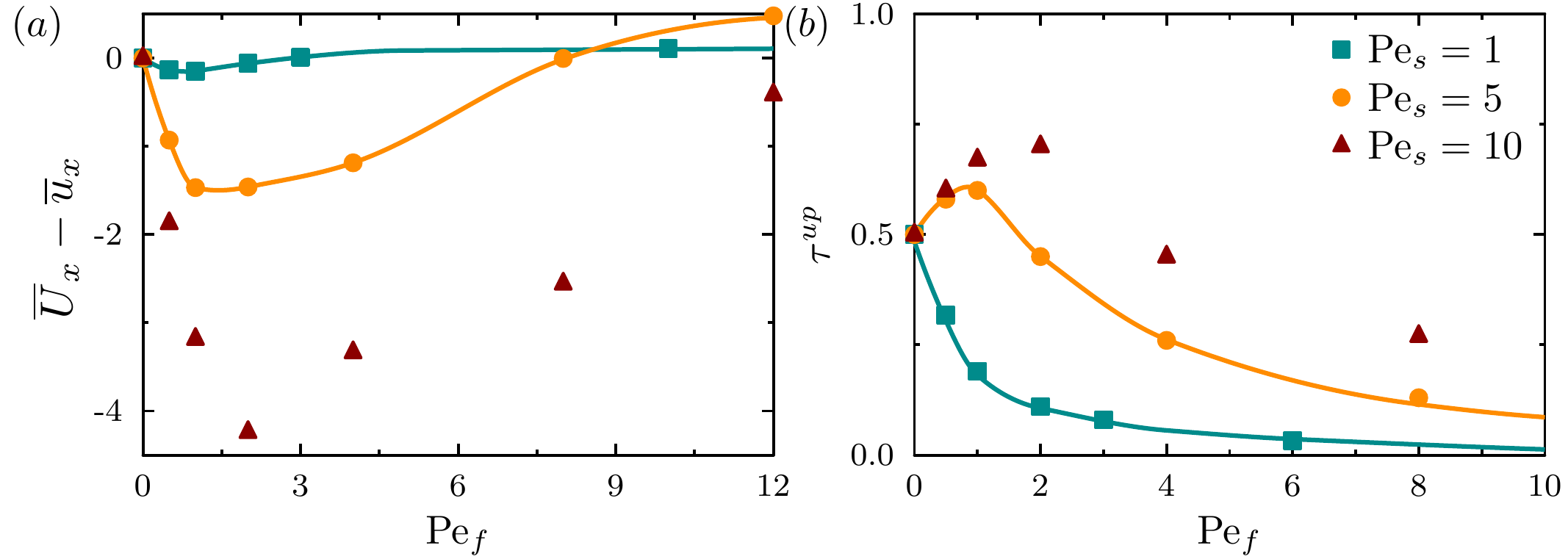}
	\caption{Mean transport in an imposed flow: ($a$) Long-time relative mean velocity $\overline{U}_x-\overline{u}_x$, where $\overline{\mathbf{U}}$ is the swimmer mean velocity and $\overline{\mathbf{u}}$ is the volume-averaged fluid velocity, as a function of $\mathrm{Pe}_f$ for different activity levels.   ($b$) Fraction of time $\tau^{up}$ during which particles are moving upstream. Results are in a square lattice with parameter values: $\kappa^2=0.1$, $\Theta_f=0$, $\epsilon_p =0.804$.\ Symbols: BD simulations; lines: analytical model.}
	\label{fig:time_up}
\end{figure}

The long-time relative velocity $\overline{U}_x-\overline{u}_x$ is plotted as a function of flow strength $\mathrm{Pe}_f$ for different values of $\mathrm{Pe}_s$ in Fig.~\ref{fig:time_up}($a$) for a square lattice of circular pillars. Here, $\overline{\mathbf{U}}$ is the mean transport velocity of the microswimmers, and $\overline{\mathbf{u}}$ is the volume-averaged fluid velocity, which is also the mean velocity that a uniformly distributed passive tracer would experience. In weak flows, active particles are found to be transported more slowly than the fluid ($ \overline{U}_x-\overline{u}_x<0$), and this tendency is especially true of fast swimmers (large $\mathrm{Pe}_s$), suggesting that a combination of wall accumulation and upstream swimming are responsible for the effect \cite{ezhilan2015channel}. In stronger flows, however, $\overline{U}_x$ increases again and in fact slightly exceeds $\overline{u}_x$ at high values of $\mathrm{Pe}_f$. To explain these trends, we consider in Fig.~\ref{fig:time_up}($b$) the fraction of time $\tau^{up}$ spent by the particles moving upstream ($\Delta \mathbf{x} \cdot \mathbf{e}_x<0$). It is directly obtained from the macrotransport theory as
\begin{equation}
\tau^{up} = \int_{S_t} \int_{\Omega} \psi_0^\infty\left(\mathbf{r},\mathbf{p}\, |\, \mathbf{J}_0^{\infty}\cdot \mathbf{{e}}_x <0\right) d\mathbf{p}\, d^2\mathbf{r}.
\end{equation}
For a purely diffusive process in the absence of any flow, the particles are equally likely to go upstream or downstream and therefore $\tau^{up}=0.5$. While weak swimmers are washed downstream for all values of $\mathrm{Pe}_f$, fast swimmers are able to overcome the flow and move upstream in weak to moderate flows. The ability of particles to re-orient upstream and swim against the flow hinges on their rotation under shear in the accumulation layers surrounding pillars; this mechanism is very similar to that responsible for upstream swimming in pressure-driven channel flows, where similar trends with respect to $\mathrm{Pe}_f$ are observed \cite{ezhilan2015channel}. Our findings are also in good qualitative agreement with recently reported experimental results for the transport of \textit{E. Coli} under strong flows \cite{creppy2018motility}, where a similar dependence of $\tau_{up}$ on flow strength was reported.


\begin{figure}[t]
	\centering
	\includegraphics[width=0.9\linewidth]{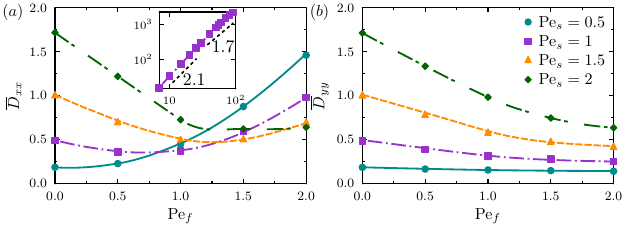}\vspace{-0.3cm}
	\caption{Dispersion in an imposed flow: ($a$)--($b$) Dependence of $\overline{D}_{xx}$ and $\overline{D}_{yy}$ on flow P\'eclet number $\mathrm{Pe}_f$ for various values of $\mathrm{Pe}_s$ in a lattice with circular pillars for an external flow in the $x$-direction ($\Theta_f=0$).\ Inset in ($a$) highlights strong-flow scalings.\ Parameter values: $\mathrm{Pe}_s=1$, $\kappa^2=0.1$, $\epsilon_p=0.804$.\ Symbols: BD simulations; lines: analytical model. } \label{fig:dispersivity} 
\end{figure}

The subtle interplay of activity and flow can be further appreciated in Fig.~\ref{fig:dispersivity}, showing the dependence of dispersion coefficients on external flow strength $\mathrm{Pe}_f$ and swimming activity $\mathrm{Pe}_s$.  In weak flows, $\overline{D}_{xx}$ and $\overline{D}_{yy}$ have similar magnitudes and the spreading process is primarily due to self-propulsion.\ As a result, faster-swimming particles (high $\mathrm{Pe}_s$) spread more efficiently by the process of active dispersion. Quite remarkably, increasing flow strength  $\mathrm{Pe}_f$ first causes a decrease in spreading.\ This decrease is unique to self-propelled particles, and indeed disappears at low values of $\mathrm{Pe}_s$. The imposed flow induces alignment of the swimmers, which prevents them from sampling all directions freely and thus inhibits spreading by active dispersion. At higher values of $\mathrm{Pe}_f$, $\overline{D}_{xx}$ starts increasing again due to shear-induced Taylor dispersion. It displays an asymptotic scaling of $\overline{D}_{xx}\sim \mathrm{Pe}_f^\gamma$ with an exponent of $\gamma\approx 2.1$ at intermediate flow strengths and of $\gamma\approx 1.7$ in very strong flows, which is the classical scaling for the porous dispersion of passive tracers \cite{edwards1991dispersion}.\ Intriguingly, activity is seen to hinder $\overline{D}_{xx}$ in strong flows in Fig.~\ref{fig:dispersivity}($a$). This stems from the fact that swimming enhances cross-stream migration, and as a result the active particle samples the imposed velocity gradient field more rapidly leading to an effective uniformity of the sensed velocity field. For a passive tracer, this effect is similar to the well-known decrease in Taylor dispersion with increasing translational diffusivity \cite{taylor1953dispersion}.\ Unlike $\overline{D}_{xx}$, the transverse dispersivity $\overline{D}_{yy}$ in Fig.~\ref{fig:dispersivity}($b$) monotonically decays with $\mathrm{Pe}_f$ as stronger flows cause faster rotations of the swimmers in the local shear and thus limit their ability to sample the $y$-direction. 

\begin{figure}[t]
	\centering
	\includegraphics[width=0.9\linewidth]{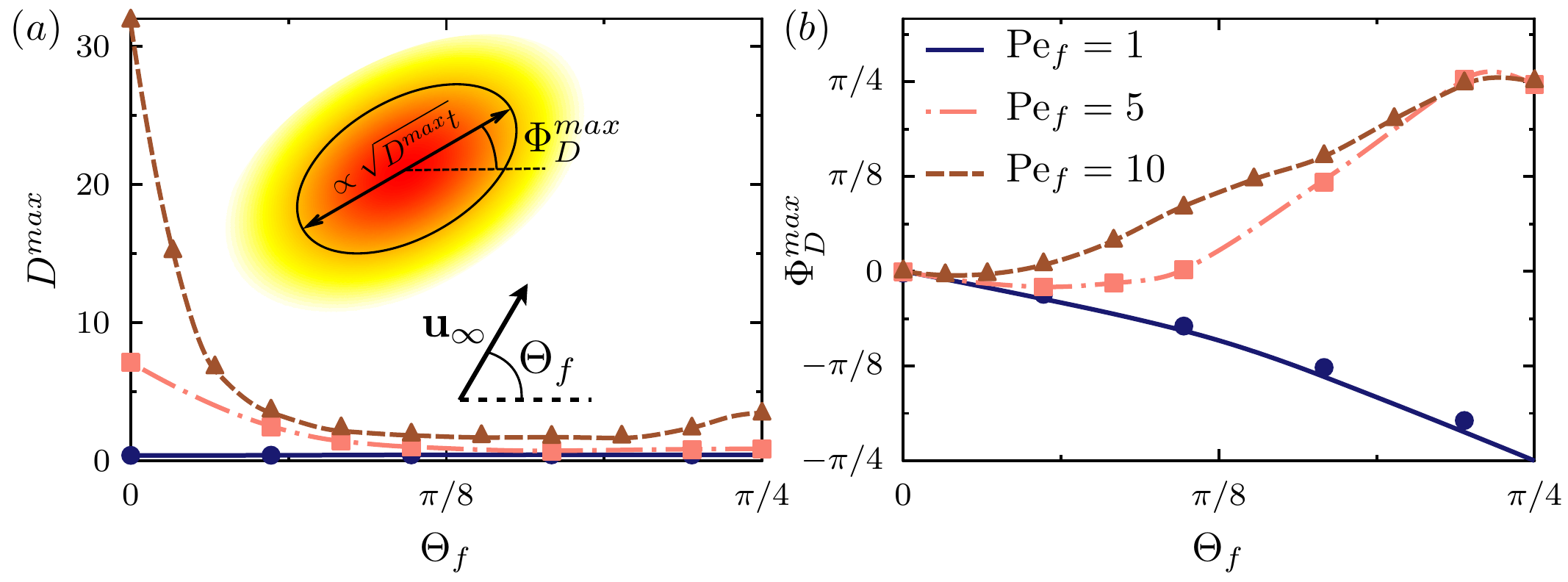}\vspace{-0.3cm}
	\caption{($a$) Maximum eigenvalue ${D}^{max}$ of the dispersivity dyadic and ($b$) corresponding direction of maximum dispersion $\Phi_D^{max}$ as a function of incoming flow angle.\ Parameter values: $\mathrm{Pe}_s=1$, $\kappa^2=0.1$, $\epsilon_p=0.804$.} \label{fig:flowangle} 
\end{figure}

Dispersion can be further controlled by varying the incoming flow angle $\Theta_f$ in Fig.~\ref{fig:flowangle}.\ For arbitrary $\Theta_f$, the dispersion tensor $\overline{\mathbf{D}}$ is no longer diagonal and maximum dispersion occurs along the eigendirection $\Phi_D^{max}$ associated with its maximum eigenvalue $D^{max}=\max (D_1,D_2)$.\ The magnitude of $D^{max}$ is maximum for $\Theta_f= 0$, which allows for easy gliding of the swimmers between rows of pillars while undergoing minimal collisions.\ Other flow directions result in more severe obstruction by the pillars and thus display much lower values of $D^{max}$, which reaches its minimum for $\Theta_f\approx \pi/8$ at high flow rates.\ While in strong flows the direction of maximum dispersion is close to that of the imposed flow, such is not the case in weak flows where activity in fact allows for significant spreading in the transverse direction as evidenced by the opposite signs for $\Theta_f$ and $\Phi_D^{max}$.

\subsection{Effect of lattice porosity, geometry and pillar shape\label{sec:latticeproperties}}

\begin{figure}[t]
	\centering
	\includegraphics[width=0.9\linewidth]{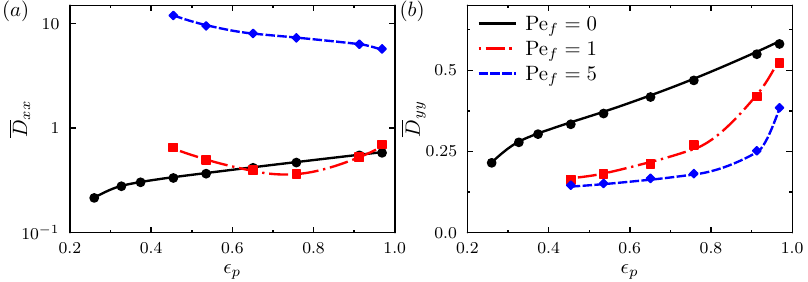}\vspace{-0.2cm}
	\caption{{Effect of lattice porosity:\ ($a$)--($b$) Dependence of $\overline{D}_{xx}$ and $\overline{D}_{yy}$ on $\epsilon_p$ in the case of circular obstacles at $\mathrm{Pe}_s=1$ and for various flow strengths $\mathrm{Pe}_f$.\ All plots are for $\Theta_f=0$ and $\kappa^2=0.1$.  }} \label{fig:latticeproperties} 
\end{figure}

The effect of lattice porosity is analyzed in Fig.~\ref{fig:latticeproperties} and further underscores the competition of active vs shear-induced dispersion. In the absence of flow and at low porosity, dispersion is weak as the pillars act as entropic barriers that strongly restrict active transport \cite{laachi2007}.\ Increasing porosity thus causes a monotonic increase in $\overline{D}_{xx}$ and $\overline{D}_{yy}$ as geometric obstruction plays a lesser role, and this dependence was indeed recently observed in experiments on \textit{Chlamydomonas reinhardtii} \cite{Brun2018}.\ In strong flows ($\mathrm{Pe}_f=5$), $\overline{D}_{xx}$ instead decreases with porosity, as smaller obstacles produce weaker shear and thus weaker dispersion.  At intermediate flow strengths ($\mathrm{Pe}_f=1$),  $\overline{D}_{xx}$ shows a non-monotonic behavior: as porosity is increased, axial dispersion first decreases due to a drop in shear rate, before increasing again at low obstacle area fractions due to active swimming.\ Curiously, the limit of $\epsilon_p\rightarrow 1$ differs from the case of swimmers in free-space, as infinitesimal obstacles still induce fluid shear in an imposed flow while also modifying the swimmer distribution in their vicinity.

\begin{figure}[t]
	\centering 
	\includegraphics[width=0.9\linewidth]{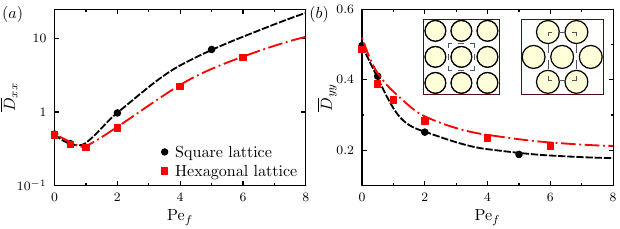}
	\caption{Variation of $\overline{D}_{xx}$ and $\overline{D}_{yy}$ as a function of $\mathrm{Pe}_f$ at a fixed porosity $\epsilon_p = 0.804$ and fixed $\mathrm{Pe}_s = 1$ for square and hexagonal lattice arrangements. For all the cases we have $\Theta_f = 0$ and $\kappa^2 = 0.1$.}
	\label{Fig:hex}
\end{figure}

The effect of lattice geometry is considered in Fig.~\ref{Fig:hex}, where we compare dispersion in a square lattice to the case of an hexagonal lattice with the same porosity. It is evident from Fig.~\ref{Fig:hex} that the hexagonal arrangement does not change any qualitative behavior. We notice that the axial dispersivity is consistently lower in a  hexagonal lattice compared to a square lattice while the behavior reverses for the transverse direction. A qualitative understanding of this behavior can be appreciated from the Lagrangian perspective of transport. In a hexagonal lattice, axial transport is hindered compared to a square lattice as straight paths between rows of obstacles are no longer available and particle trajectories must curve around between staggered pillars. These curved paths are accompanied by more frequent collisions with pillars, which also has the effect of enhancing transverse motion, thus explaining the increase in $\overline{D}_{yy}$. 

\begin{figure}[t]
	\centering\vspace{0.0cm}
	\includegraphics[width=0.9\linewidth]{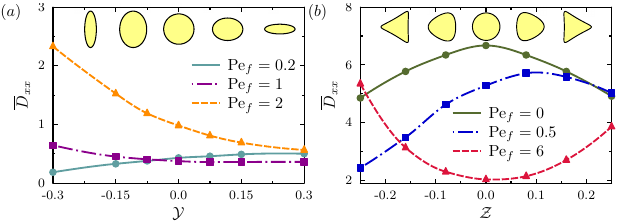}\vspace{-0.0cm}
	\caption{{Effect of obstacle geometry:\ ($a$)--($b$) Effect of shape aspect ratio parameter $\mathcal{Y}$  and fore-aft asymmetry $\mathcal{Z}$ on $\overline{D}_{xx}$ at $\mathrm{Pe}_s=4$ for different values of $\mathrm{Pe}_f$.\ Inset shows obstacle shapes.\ All plots are for $\Theta_f=0$ and $\kappa^2=0.1$.  }} \label{fig:pillarshapes} 
\end{figure}

Pillar aspect ratio and fore-aft asymmetry both also affect transport in non-trivial ways. As shown in Fig.~\ref{fig:pillarshapes}($a$), elliptical obstacles aligned with a weak flow cause higher streamwise dispersion than if aligned perpendicular to it, as the former allow swimmers to glide past while the latter act as entropic barriers that obstruct transport \cite{laachi2007}.\ These trends reverse at high flow rates, as perpendicular obstacles induce stronger velocity gradients thereby enhancing shear-induced dispersion.
Fore-aft asymmetric obstacles also lead to complex trends in Fig.~\ref{fig:pillarshapes}($b$).\ In the absence of flow, $\overline{D}_{xx}$ decreases as the shape parameter $\mathcal{Z}$ deviates from $0$ (circle) by the aforementioned entropic obstruction mechanism. When a flow is applied, fluid shear causes particles to swim along the walls and against the flow, resulting in an enhancement of particle accumulation around obstacles for $\mathcal{Z}<0$ and in a reduction for $\mathcal{Z}>0$.\ Consequently, the effects of shear-alignment (in weak flows) and shear-induced dispersion (in strong flows) on transport are magnified for obstacle shapes that promote wall accumulation ($\mathcal{Z}<0$), as seen in Fig.~\ref{fig:latticeproperties}($d$).\ This asymmetry becomes negligible at very high values of $\mathrm{Pe}_f$, as activity becomes subdominant and the magnitude of the shear gradients that drive dispersion in this limit is independent of the sign of $\mathcal{Z}$ by reversibility of Stokes flow.

\begin{figure}[t]
	\centering\vspace{0.0cm}
	\includegraphics[width=0.47\linewidth]{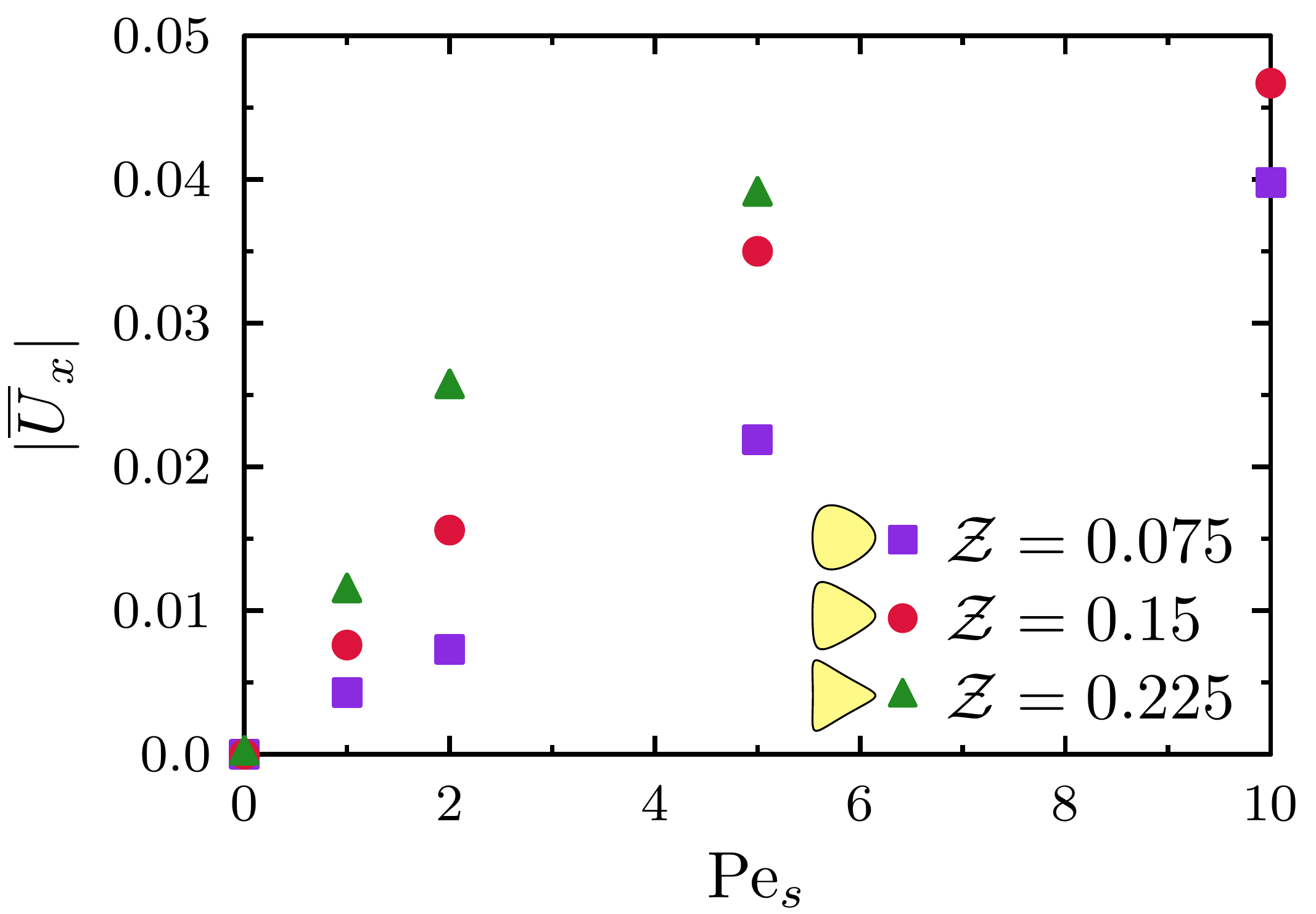}\vspace{-0.1cm}
	\caption{{Mean transport velocity $\overline{U}_x$ in square arrays of fore-aft asymmetric pillars (asymmetry parameter $\mathcal{Z}$) as a function of swimming strength $\mathrm{Pe}_s$ in the absence of flow ($\mathrm{Pe}_f=0$). Results are from BD simulations with $\Theta_f=0$, $\kappa^2=0.1$ and $\mathcal{Y}=0$.  }} \label{fig:shapevelocity} 
\end{figure}

In addition to their effect on dispersion, fore-aft asymmetric pillars can also introduce a net bias in the mean transport, which is most appreciable in the absence of flow ($\mathrm{Pe}_f=0$) where a non-zero transport velocity is observed. This is illustrated in Fig.~\ref{fig:shapevelocity}, where increasing either $\mathcal{Z}$ or $\mathrm{Pe}_s$ amplifies the effect. This net migration, which results from irreversible collisions between the active particle and the curved boundary, occurs towards the pointed end of the pillar in agreement with experiments on catalytic rods in teardrop arrays \cite{wykes2017guiding} and a related statistical model \cite{tong2017directed}. 

\section{Concluding remarks\label{sec:conclusions}}

We have analyzed the long-time transport properties of active particles in a porous lattice under both quiescent and flow conditions. We developed a continuum model based on generalized Taylor dispersion theory, and started from a single-particle level approach to show that the overall behavior of a dilute cloud of cells can be described by an obstacle-free advection-diffusion equation, whose effective long-time mean particle velocity and dispersivity dyadic have been determined through a set of boundary value problems. The predictions of our continuum model, which agree perfectly with Brownian dynamics simulations at long times, highlight the complex interplay of particle motility, alignment under shear, cross-stream migration, lattice porosity and pillar geometry on asymptotic dispersion, and we have provided a physical explanation for the predicted trends. In particular, we showed that obstacles behave predominantly as entropic barriers at low flow rates and as regions of shear production at high flow rates, and found that shear-induced polarization as well as activity-driven cross-stream migration play an important role on active particle dispersion. \ Our theory provides a simple framework for analyzing microorganismal transport in natural structured environments and for designing engineered porous media in applications involving microswimmers.\ The fundamental premise of non-interacting active Brownian particles glosses over many details that may become relevant in specific systems.\ For instance, future work should address the roles of hydrodynamic interactions with obstacles \cite{spagnolie2015geometric}, swimmer-specific scattering dynamics \cite{kantsler2013scatter}, rheological effects due to activity \cite{saintillan2018rheology}, and the potential emergence of spontaneous flows and active turbulence in denser systems \cite{theillard2017geometric,theillardjcp}.

\begin{acknowledgements}
D.S. gratefully acknowledges funding from NSF Grant DMS-1463965.
\end{acknowledgements}

\appendix

\section{Conformal mapping for non-circular obstacles}\label{app:CM}

In order to explore the effect of pillar shape on active particle transport, we consider a class of non-circular pillar geometries that are obtained using conformal mappings. Specifically, we make use of  following Riemann map of the unit disc:
\begin{equation}
z(\sigma) = \mathcal{W} \sigma + \frac{\mathcal{Y}}{\sigma} + \frac{\mathcal{Z}}{\sqrt{2}\, \sigma^2},
\end{equation}
where $\sigma = \mathrm{e}^{\mathrm{i} \chi}$ with polar angle $\chi \in [0,2\pi)$. $\{\mathcal{W,Y,Z}\}$ are three parameters that are varied to change the obstacle shape. An identical conformal map was previously used to study amoeboid swimming \cite{avron2004optimal}. Avoiding self-intersection of the boundary provides the following constraint on the shape parameters \cite{avron2004optimal}:
\begin{equation}
\mathcal{W} \ge \mathcal{Y} \pm \sqrt{2}\, \mathcal{Z}, \hspace*{4mm} \mathcal{W Y} \ge 2 \mathcal{Z}^2 - \mathcal{W}^2.
\end{equation}
For a given lattice porosity we also need to constrain the area of these shapes, which can be shown  to be \cite{avron2004optimal}
\begin{equation}
A = \pi \left(\mathcal{W}^2 - \mathcal{Y}^2 -\mathcal{Z}^2\right).
\end{equation}
Following our non-dimensionalization, we choose $A$ to be the area of a circle with unit radius. Given the value of $A$, we need only vary parameters  $\{\mathcal{Y},\mathcal{Z}\}$, which control shape aspect ratio and fore-aft asymmetry, respectively. Typical shapes obtained by this method are illustrated in Figs.~\ref{fig:geo} and \ref{fig:pillarshapes}.

\section{Flow calculation: boundary integral method}\label{app:BI}

We use the boundary integral method \cite{pozrikidis92} to calculate the fluid velocity. Owing to the periodicity of the lattice, we can make use of the doubly periodic Green's function for two-dimensional Stokes flow, which is computed using standard Ewald summation techniques. The fluid velocity $\mathbf{u}$ at any point inside the unit cell of the lattice is decomposed into the free-stream component $\mathbf{u}^\infty$ and a disturbance velocity $\mathbf{u}^D$ generated due to the presence of the obstacles. In the present problem we  take $\mathbf{u}^\infty$ to be a uniform streaming flow. On the surface $\partial C_w$ of the obstacles, the no-slip condition $\mathbf{u} = \mathbf{0}$ is satisfied, or alternately, $\mathbf{u}^D = -\mathbf{u}^\infty$. We use a single-layer representation for the velocity to obtain:
\begin{equation}\label{bem}
\mathbf{u}(\bx_0) = \mathbf{u}^\infty(\bx_0) - \frac{1}{4 \pi \mu} \sum_{q=0}^{N_p} \int_{C_q} \mathbf{G} (\bx,\bx_0)\cdot \mathbf{f}(\bx)\, d\ell(\bx),
\end{equation}
where $N_p$ is the total number of obstacles within the unit cell, $C_q$ is the contour of the $q^{th}$ obstacle, $\mathbf{G}(\bx,\bx_0)$ is the computed Green's function, and $\mathbf{f}(\bx)$ is the local traction. On enforcing the no-slip boundary condition one is able to solve for the unknown tractions by inverting a dense linear system. Once the tractions are obtained, the fluid velocity can be calculated at any point using Eq.~\eqref{bem}. For all the cases presented in the paper we have used linear elements to discretize the contour of the obstacles and Gauss-Legendre quadrature for evaluating the integrals. Sangani and Acrivos \cite{sangani1982slow} studied the flow past a periodic array of circular cylinders using the streamfunction-vorticity formulation and provided results for the drag on a cylinder as a function of obstacle volume fraction: our numerical method was validated against their predictions and excellent agreement was found.

In order to facilitate fast computation, the velocity was pre-tabulated on a Cartesian mesh and bilinear interpolation was used subsequently. We used second-order-accurate finite-difference approximations to compute and tabulate velocity gradients on the mesh. Our interpolation results were validated against the boundary integral method and the error was always within one percent.

\section{Brownian dynamics algorithm}\label{app:BD}

We performed Brownian dynamics simulations using discrete point swimmers, whose trajectories follow the Langevin equations (\ref{eq:xdot})--(\ref{eq:pdot}). We parametrize the director as $\mathbf{p}=(\cos\theta,\sin\theta)$. During time step $\Delta t$ in a simulation, the changes in position $\mathbf{R}$ and orientation $\theta$ are calculated as
\begin{align}
\Delta\mathbf{R} &= [\mathrm{Pe}_s \bp +\mathrm{Pe}_f \bu]\,\Delta t+ \sqrt{2 \kappa^2 \Delta t}\, \boldsymbol{ \zeta}_t, \\
\Delta {\theta} &=  \mathrm{Pe}_f[\bp^\perp \cdot \nabla\mathbf{u}^T \cdot \bp]\,\Delta t   + \sqrt{2\Delta t}\, \zeta_r,
\end{align}
where the equations have been made dimensionless. Here, $\mathbf{p}^\perp=(-\sin\theta,\cos\theta)$, and 
$\boldsymbol{\zeta}_t$ and $\zeta_r$ are Gaussian random variables with zero mean and unit variance. In most simulations we use $N = 10^5$ particles. The no-flux boundary condition on the surface of the obstacles is enforced using a reflecting condition. The algorithm was validated against well-known results for the transport of passive tracers in porous media \cite{edwards1991dispersion}.

\bibliography{bibfile}

\end{document}